\documentclass[11pt,a4paper,oneside]{article}
\usepackage{graphicx,amssymb,amsmath}
\usepackage[linkcolor={blue},citecolor={red},colorlinks=true]{hyperref}
\usepackage{cite}
\usepackage{relsize}
\usepackage{enumerate}
\usepackage[margin=3 cm]{geometry}
\usepackage{slashed}
\usepackage{multirow}
\usepackage[normalem]{ulem}
\usepackage[dvipsnames, x11names, table]{xcolor}
\usepackage{float}
\usepackage{cancel}

\usepackage[font=small,labelfont=bf]{caption}
\usepackage[font=small,labelfont=bf]{subcaption}
\usepackage{siunitx}
\usepackage{gensymb}
\usepackage{graphicx}
\usepackage{wrapfig}
\usepackage{chngcntr}
\counterwithout{figure}{section}
\counterwithout{table}{section}
\usepackage[font=small]{caption}

\usepackage{soul}

\numberwithin{equation}{section}

\def\beq{\begin{equation}}
\def\eeq{\end{equation}}
\def\bea{\begin{eqnarray}}
\def\eea{\end{eqnarray}}

\def\bit{\begin{itemize}}
\def\eit{\end{itemize}}

\def\baa{\begin{array}}
\def\eaa{\end{array}}

\def\simgt{\mathrel{\lower2.5pt\vbox{\lineskip=0pt\baselineskip=0pt
           \hbox{$>$}\hbox{$\sim$}}}}
\def\simlt{\mathrel{\lower2.5pt\vbox{\lineskip=0pt\baselineskip=0pt
           \hbox{$<$}\hbox{$\sim$}}}}

\def\bfc{\begin{figure}\begin{center}}
\def\efc{\end{center}\end{figure}}

\definecolor{chromeyellow}{rgb}{1.0, 0.65, 0.0}
\definecolor{darkcoral}{rgb}{0.8, 0.36, 0.27}
\definecolor{cadmiumgreen}{rgb}{0.0, 0.42, 0.24}

\providecommand{\abs}[1]{\lvert#1\rvert}
\providecommand{\bd}[1]{\boldsymbol{#1}}
\providecommand{\ro}[1]{\mathrm{#1}}

\setstcolor{blue}
\sethlcolor{cyan}

\makeindex

\begin{document}

\begin{center}

\hspace{-0.4cm}{\Large \bf 
Anharmonic Effects on the Squeezing of Axion Perturbations\\}

\vspace{1cm}{Valentina Danieli$^{a,b,c,1}$, Takeshi Kobayashi$^{a,b,c,d,2}$, Nicola Bartolo$^{e,f,g,3}$, Sabino Matarrese$^{e,f,g,h,4}$, Matteo Viel$^{a,b,c,i,j,5}$}
\\[7mm]
 {\it \small

$^a$ SISSA International School for Advanced Studies, Via Bonomea 265, 34136, Trieste, Italy\\[0.15cm]
$^b$ INFN - Sezione di Trieste, via Valerio 2, 34127, Trieste, Italy\\[0.1cm]
$^c$ IFPU, Institute for Fundamental Physics of the Universe, Via Beirut 2, 34014 Trieste, Italy\\[0.1cm]
$^d$ Kobayashi-Maskawa Institute for the Origin of Particles and the Universe, Nagoya University, Nagoya 464-8602, Japan\\[0.1cm]
$^e$ Dipartimento di Fisica e Astronomia ``G. Galilei'', Universit\`{a} degl Studi di Padova,  
Via Francesco Marzolo 8, 35131, Padova, Italy \\[0.1cm] 
$^f$ INFN - Sezione di Padova, Via Francesco Marzolo 8, 35131, Padova, Italy\\[0.1cm]
$^g$ INAF - Osservatorio Astronomico di Padova, Vicolo dell’Osservatorio 5, 35122 Padova, Italy\\[0.1cm]
$^h$ Gran Sasso Science Institute, Viale F. Crispi 7, I-67100 L’Aquila, Italy\\[0.1cm]
$^i$ INAF - Osservatorio Astronomico di Trieste, Via G.B. Tiepolo 11, I-34131 Trieste, Italy\\[0.1cm]
$^j$ ICSC - Italian Research Center on High Performance Computing, Big Data and Quantum Computing, Italy

}
\end{center}

\bigskip \bigskip \bigskip

\centerline{\bf Abstract} 
\begin{quote}
It is assumed in standard cosmology that the Universe underwent a period of inflation in its earliest phase, providing the seeds for structure formation through vacuum fluctuations of the inflaton scalar field. These fluctuations get stretched by the quasi-exponential expansion of the Universe and become squeezed.
The aim of this paper is to deepen the understanding of the squeezing process, considering the effect of self-interactions.
Axion-like particles can provide a useful setup to study this effect.
Specifically we focus on the consequences that a non-trivial evolution of the background axion field has on the squeezing of the perturbations.
We follow the evolution of the axion's fluctuation modes from  the horizon exit during inflation to the radiation-dominated epoch.
We compute Bogoliubov coefficients and squeezing parameters, which are linked to the axion particle number and isocurvature perturbation. We find that the quantum mechanical particle production and the squeezing of the perturbations are enhanced, if one accounts for anharmonic effects, i.e., the effect of higher order terms in the potential. This effect becomes particularly strong towards the hilltop of the potential.
\end{quote}

\vfill
\noindent\line(1,0){188}
{\scriptsize{ \\ E-mail:
\texttt{$^1$\href{vdanieli@NOSPAMsissa.it}{vdanieli@sissa.it}},
\texttt{$^2$\href{takeshi.kobayashi@NOSPAMsissa.it}{takeshi.kobayashi@sissa.it}},

\texttt{$^3$\href{mailto:nicola.bartolo@pd.infn.it}{nicola.bartolo@pd.infn.it}},
\texttt{$^4$\href{sabino.matarrese@pd.infn.it}{sabino.matarrese@pd.infn.it}},
\texttt{$^5$\href{viel@NOSPAMsissa.it}{viel@sissa.it}}
}}

\newpage

\newpage

\tableofcontents

\newpage

\section{Introduction}

The current version of the standard cosmological model assumes that the Universe underwent a period of quasi-exponential expansion in its earliest stage, which solves several puzzles of the Big Bang cosmological scenario. This period of inflation turned out to be also an excellent mechanism to explain the origin of structure formation in the Universe. The exponential stretching of the fluctuations of the scalar field driving inflation, the so-called inflaton field, provides the seeds for the later gravitational collapse. In most inflationary models, these fluctuations have a quantum origin, however the CMB sky we observe today is classical. Therefore the question about the quantum or classical origin of the initial perturbations and how to probe it arise.

The classicalization of a quantum perturbation has been a subject of active research (see for example \cite{Polarski:1995jg, Grishchuk:1990bj, Grishchuk:1997nr, Grishchuk:2005qe, Kiefer:1998jk,Allen:1999xw}). Actually, inflation itself provides an explanation for the ``classicalization'' of the originally quantum perturbations: they are squeezed due to the fast expansion of the Universe. A squeezed state is a special quantum state for which one variable has an arbitrarily small uncertainty, while its conjugate counterpart has a very big uncertainty, correspondingly. Since squeezed states are associated with a huge uncertainty in one variable, they are highly quantum mechanical states \cite{Hsiang:2021kgh}. However, within standard cosmological measurements, a squeezed state is indistinguishable from a classical phase-space distribution, if one considers a Gaussian state \cite{Martin:2015qta}. Most of the previous works have focused on the study of nearly free fields. The aim of this paper is to deepen the understanding of the squeezing process considering the effect of self-interactions.

In general, to address the problem of the nature of the perturbations there is one further mechanism that takes place in the Early Universe that has to be taken into account: the decoherence during reheating. In fact, in order for a squeezed fluctuation to maintain its quantum nature into the later Universe, it is necessary that it does not undergo decoherence through the interactions with the cosmological plasma. The phenomenon of decoherence has been widely discussed in the literature; see for example \cite{Lombardo:2005iz,Martineau:2006ki,Campo:2008ju,Campo:2008ij,Burgess:2006jn}.

From this perspective, axion-like particles~\cite{Ringwald:2012hr,Marsh:2017hbv} coupled to other matter fields only through gravitational interactions provide an ideal setup that can evade decoherence. In particular, axions can be produced non-thermally from an initial displacement of the axion field, which is called the misalignment mechanism. In this model axion-like particles are not required to directly interact with the thermal bath.
Another interesting feature of axion-like particles is that their fluctuations around the classical axion background source axion isocurvature perturbations, that can be directly linked with observables. Their correlators constitute in principle a promising way to investigate the quantum origin of the cosmological perturbations. Indeed axions may evade decoherence and their bispectrum could provide a way to discern among the quantum or classical nature of the perturbations (see \cite{Green:2020whw, Martin:2018lin, DaddiHammou:2022itk}).
One further advantage in working with axions is that several constraints are already present in the literature, giving rise to strong bounds on the cosmological observables.
Also for the purpose of studying interacting fields, axions are useful since they necessarily have self-interactions due to their periodic potential. The effects induced by higher-order terms in the potential is often referred to as ``anharmonic effects,'' and axions provide a computable setup for incorporating the effect of self-interactions. Indeed in the late Universe the axion field settles down to the minimum of its potential and thus the self-interaction switches off. Axions thus reduce to effectively free fields in the asymptotic future, which allows one to apply Bogoliubov computations for studying the evolution of the fluctuations.

In this work we study the effect of self-interactions on the squeezing of axion fluctuations, focusing on how the background evolution modifies the evolution of the perturbations themselves. We focus on axion-like particles (instead of QCD axions~\cite{Peccei:1977hh,Weinberg:1977ma,Wilczek:1977pj}), and  consider a scenario where the Peccei-Quinn symmetry is broken during inflation. In this way the axion field gets homogenized, giving rise to a background axion condensate. 
We study the perturbations around this background. In previous works on axion squeezing \cite{Kuss:2021gig,Kanno:2021vwu}, only the quadratic term has been considered.
In this paper we go beyond these analyses, considering the effect of self-interactions. In particular we study the interaction between the perturbations and the zero-mode, i.e. the background axion field, neglecting the interactions among perturbations themselves. We find that the interaction with a non-trivial background gives already interesting effects. In particular, it strongly enhances the squeezing of the perturbations, which is directly linked with the number of particles created due to the expansion of the background.
It has been known that anharmonic effects in the misalignment scenario give rise to the enhancement of the axion density~\cite{Turner:1985si,Bae:2008ue}, as well as the isocurvature perturbation~\cite{Lyth:1991ub,Strobl:1994wk,Kobayashi:2013nva}.
We revisit this effect using Bogoliubov calculations, and shed light on the anharmonic enhancement on the axion squeezing.

The paper is organised as follows. In Section \ref{axiontheory} we present our setup, and introduce the concepts of Bogoliubov coefficients and squeezing parameters. In Section \ref{background} the evolution in time of the background field is analysed. In Section \ref{axionfluct} the numerical results for the mode functions, the Bogoliubov coefficients and the squeezing parameters are reported. Their interpretation in terms of cosmological particle creation and squeezing are also presented. Moreover the dependence of the physical observables on the parameters of the system is studied. We conclude and discuss our results in Section \ref{conclusions}. In Appendix~\ref{sqparphysmean} we give explanations on the squeezing parameters, while in Appendix \ref{quadratic} we provide analytic calculations for a quadratic axion potential. Finally, in Appendix \ref{morerealistic} the cases of a non-instantaneous reheating and a quasi-de Sitter spacetime are considered.

\section{Quantization of Axion Fluctuations}
\label{axiontheory}

\subsection{Setup}
\label{backmodel}

We fix the metric to a flat FRW, $ds^2 = a^2(\tau)\,\left( -d\tau^2 + d\boldsymbol{x}^2 \right)$. Here $\tau$ is the conformal time, linked with the cosmic time via $dt=a(\tau)\,d\tau$. We consider the evolution of the Universe as simply given by a de Sitter (dS) phase of inflation, followed by a period of radiation domination (RD). We further assume, for simplicity, that the reheating phase is instantaneous\footnote{The cases of non-instantaneous reheating and quasi-de Sitter inflation are considered in Appendix~\ref{morerealistic}.}. In the simple scenario here introduced, the Hubble rate is given by:

\begin{equation}
\label{Hubbletau}
H = 
\begin{cases}
    H_{\ro{inf}} & \qquad \text{\small{de Sitter}} \\
    H_{\ro{inf}}\, \left(a/a_\ro{reh}\right)^{-2} & \qquad \text{\small{Radiation Domination}}
\end{cases}
\end{equation}
Here $H_{\ro{inf}}$ is the constant Hubble scale during the de Sitter phase and $a_\ro{reh}$ is the value of the scale factor at the reheating.
We consider the axion to have negligible effect on the cosmological evolution. It is assumed that the Peccei-Quinn symmetry is broken before inflation, and specifically that $f > \text{max}(T_{\ro{reh}}, H_{\ro{inf}})$, where $f$ is the symmetry breaking scale and $T_{\ro{reh}}$ is the reheating temperature. This ensures that the symmetry is not restored during the reheating phase.
For our assumption of instantaneous reheating, $T_\ro{reh} \sim \sqrt{M_{\ro{Pl}} H_\ro{inf}} > H_\ro{inf}$, hence it is enough to require $f > T_{\ro{reh}}$.

Let us now consider the axion potential, which can be written as
\begin{equation}
\label{V}
    V(\phi) = f^2\, m^2 \left[ 1-\cos\left( \frac{\phi}{f} \right) \right]\,,
\end{equation}
where $\phi$ is the axion field. 
We assume that the axion mass is constant in time, and also that $m<H_{\ro{inf}}$; this ensures the axion mass becomes comparable with the Hubble rate during radiation domination, after reheating has occurred. After the time when $m\sim H_\ro{inf}$, the field starts oscillating around its minimum and the equations of motion becomes that of a damped oscillator.

We consider the axion field to be a homogeneous background plus vacuum fluctuations. The initial value of the axion field affects the subsequent evolution of the background. Specifically, if the starting point is near the hilltop of the potential then the field scans the full cosine potential as it rolls down.
When deviations from the quadratic potential are considered, the perturbations are affected by the background evolution.
These perturbation modes exit the horizon during inflation. After inflation has ended they re-enter the horizon. Small-scale modes re-enter the horizon before the axion field begins to oscillate;
hence they propagate unaffected by the oscillations. On the contrary, large-scale modes feel the oscillations and their evolution in time is consequently affected. Since we focus on the effect of the oscillations on the evolution of the perturbations, we will focus on large-scale modes that re-enter the horizon after the onset of the oscillations.

\subsection{Canonical Quantization}
\label{canoquant}

In order to compute the squeezing parameters describing the system, we first need to calculate the Bogoliubov coefficients for the axion field.
We work with the full potential for the background field evolution, while 
considering only the first order expansion of the potential for the perturbations. The equation of motion for the axion field is:
\begin{equation}
\label{phi}
    \ddot{\phi} + 3H\dot{\phi} - \frac{1}{a^2} \partial_i^2 \phi + \frac{dV(\phi)}{d\phi}=0\,.
\end{equation}
Here the dot indicates a derivative with respect to the physical time $t$, and Latin letters are used to denote spatial indices. We split the axion field as 
\begin{equation}
 \phi (t, \bd{x}) = \bar{\phi} (t) + \delta \phi (t, \bd{x})\,,
\label{eq:split}
\end{equation}
where $\bar \phi$ is the homogeneous classical background field, and $\delta\phi$ is the small perturbation around it. We are going to expand the axion up to quadratic order in $\delta \phi$. The first derivative of the potential, up to linear order, can be written as
\begin{equation}
\label{V2}
\begin{split}
    V^\prime(\phi) = &f m^2 \left[ \sin \left(\frac{\bar \phi}{f}\right)\, + \cos \left(\frac{\bar \phi}{f}\right)\, \left(\frac{\delta \phi}{f}\right) \right]\,.
\end{split}
\end{equation}
The equation of motion for the homogeneous background field becomes
\begin{equation}
\label{phi0}
    \ddot{\bar\phi} + 3H \dot{\bar\phi} + f m^2 \sin \left(\frac{\bar \phi}{f}\right) = 0\,,
\end{equation}
while at first order in the perturbations one has
\begin{equation}
\label{deltaphi}
    \delta\ddot{\phi}_{\boldsymbol{k}} + 3H\delta \dot{\phi}_{\boldsymbol{k}} + \left[ \frac{k^2}{a^2} + m^2 \cos \left(\frac{\bar \phi}{f}\right) \right] \delta \phi_{\boldsymbol{k}}=0\,.
\end{equation}
Here we have moved to Fourier space,
\begin{equation}
   \delta \phi_{\bd{k}} = \int \frac{d^3 x}{(2\pi)^{3/2}}
e^{-i\, \boldsymbol{k} \cdot \boldsymbol{x}} \, \delta \phi,
\end{equation}
and we used $k=|\bd{k}|$. The corresponding quadratic action for the perturbations is
\begin{equation}
\label{S}
    S = \int d^3x d\tau \, a^2 \left[ \frac{1}{2}\delta\phi^{\prime 2} - \frac{1}{2} (\partial_i\delta\phi)^2 - \frac{1}{2} m^2a^2 \cos \left(\frac{\bar \phi}{f}\right)\, \delta\phi^2 \right]\,.
\end{equation}
Here the prime denotes a $\tau$-derivative.
One can now introduce a new variable $\chi=a\delta\phi$. Then the action can be rewritten as
\begin{equation}
\label{S1}
    S = \int d^3x d\tau\, \frac{1}{2}\, \left[ \chi^{\prime\,2} + \left(\frac{a^\prime}{a} \right)^2 \chi^2 - 2\frac{a^\prime}{a}\, \chi\chi^\prime -\left(\partial_i\chi\right)^2- m^2a^2 \cos \left(\frac{\bar \phi}{f}\right)\, \chi^2 \right]\,.
\end{equation}
The mixed term $\chi\chi^\prime$ can be removed by integrating by parts and dropping surface terms as
\begin{equation}
\label{Smass2}
    S = \int d^3x d\tau\, \frac{1}{2}\, \left[ \chi^{\prime\,2} - (\partial_i \chi)^2 - \left( m^2a^2\,\cos\left(\frac{\bar \phi}{f}\right) - \frac{a^{\prime\prime}}{a} \right)\chi^2 \right]\,.
\end{equation}
From this action we can construct the Hamiltonian as $\mathcal{H} = \int d^3 x \left( p \chi' - \mathcal{L} \right)$, where $p = \chi^\prime$ and $\mathcal{L}$ is the Lagrangian. Writing in terms of the Fourier components, the Hamiltonian acquires the following form:
\begin{equation}
\label{Hmass}
    \mathcal{H} = \frac{1}{2}\,\int d^3k\, \left[ p_{\boldsymbol{k}}p^*_{\boldsymbol{k}} + \left( k^2 + m^2_{\mathrm{eff}}\,a^2 - \frac{a^{\prime\prime}}{a} \right)\,\chi_{\boldsymbol{k}}\chi^*_{\boldsymbol{k}} \right]\,,
\end{equation}
where the effective mass $m^2_{\mathrm{eff}}$ is given by
\begin{equation}
\label{meff}
    m^2_{\mathrm{eff}} = m^2\, \cos \left(\frac{\bar \phi}{f}\right)\,.
\end{equation}
Notice that the effective mass depends on time through the evolution of the background~$\bar{\phi}$. Moreover it can become tachyonic, i.e. $m^2_{\mathrm{eff}} < 0$, due to the cosine factor. We also introduce the time dependent frequency $\omega_k$, defined as
\begin{equation}
\label{omega}
    \omega^2_k = k^2 + m^2_{\mathrm{eff}}a^2 - \frac{a''}{a}\,.
\end{equation}

In order to quantize the fields, we introduce time-dependent ladder operators as follows:
\begin{equation}
\label{u}
\chi_{\boldsymbol{k}}(\tau) = \frac{1}{\sqrt{2|\omega_k(\tau)|}} \left( a_{\boldsymbol{k}}(\tau) + a^\dagger_{-\boldsymbol{k}}(\tau) \right)\,,
\end{equation}
\begin{equation}
\label{p}
p_{\boldsymbol{k}}(\tau) = -i \sqrt{\frac{|\omega_k(\tau)|}{2}} \left( a_{\boldsymbol{k}}(\tau) - a^\dagger_{-\boldsymbol{k}} (\tau) \right)\,,
\end{equation}
respecting the canonical commutation relations\footnote{In Fourier space the commutation relation between the field $\chi_{\boldsymbol{k}}(\tau)$ and its conjugate momentum $p_{\boldsymbol{k}^\prime}(\tau)$ is given by: $\left[ \chi_{\boldsymbol{k}}(\tau), p^\dagger_{\boldsymbol{k}^\prime}(\tau) \right] = i \delta^{(3)}(\boldsymbol{k}-\boldsymbol{k}^\prime)$.}
\begin{equation}
\label{commu}
    \left[ \chi(\tau, \boldsymbol{x}), p (\tau, \boldsymbol{x}^\prime) \right] = i \delta^{(3)}(\boldsymbol{x}-\boldsymbol{x}^\prime)\,,\qquad \qquad \left[ a_{\boldsymbol{k}}(\tau), a^\dagger_{\boldsymbol{k}^\prime}(\tau) \right] = \delta^{(3)}(\boldsymbol{k}-\boldsymbol{k}^\prime)\,.
\end{equation}
All the other commutators vanish. One should note that the frequency $\omega_k$ can become imaginary, hence we chose to normalize the fields with $|\omega_k|$ in (\ref{u}) and (\ref{p}).

\subsection{Bogoliubov Transformation}
\label{bogotransformation}

Let us introduce an `initial' time $\tau_0$ during inflation when the wave modes of interest are still deep inside the Hubble horizon. 
Then the time-dependent ladder operators at later times are related to those at $\tau_0$ through a Bogoliubov transformation,
\begin{equation}
\label{bogo}
\begin{split}
 &a_{\boldsymbol{k}}(\tau) = \alpha_k(\tau)\, a^0_{\boldsymbol{k}} + \beta_k(\tau) a^{0\,\dagger}_{-\boldsymbol{k}}\, ,
\\
 &a^\dagger_{-\boldsymbol{k}}(\tau) = \alpha^*_k(\tau)\, a^{0\,\dagger}_{-\boldsymbol{k}} + \beta^*_k(\tau) a^0_{\boldsymbol{k}}\, ,
\end{split}
\end{equation}
where $a^0_{\boldsymbol{k}} = a_{\boldsymbol{k}}(\tau_0)$.
From the commutation relations for the ladder operators, one finds the following relation among the Bogoliubov coefficients:
\begin{equation}
\label{commu1}
    \lvert \alpha_k \rvert ^2 - \lvert \beta_k \rvert ^2 = 1\,.
\end{equation}

The variables $\chi_{\boldsymbol{k}}$ and $p_{\boldsymbol{k}}$ can be written in terms of the time-independent ladder operators directly, introducing the mode  functions $u_k(\tau)$:
\begin{equation}
\label{umass}
    \chi_{\boldsymbol{k}} = u_k \, a^0_{\boldsymbol{k}} + u^*_k \, a^{0\,\dagger}_{-\boldsymbol{k}}\,,
\end{equation}
\begin{equation}
\label{pmass}
    p_{\boldsymbol{k}} = u^\prime_k \, a^0_{\boldsymbol{k}} + u^{*\,\prime}_k \, a^{0\,\dagger}_{-\boldsymbol{k}}\,.
\end{equation}
The normalization condition for the mode functions is given by
\begin{equation}
\label{normalization}
u_k\,u^{*\,\prime}_k - u^*_k\,u^{\prime}_k = i\,.
\end{equation}
It arises from the commutations of the real space variables and the ladder operators.
The mode functions can then be used to find an expression for the Bogoliubov coefficients $\alpha_k$ and $\beta_k$. Indeed, rewriting (\ref{u}) and (\ref{p}) one gets:
\begin{equation}
\begin{split}
 &a_{\boldsymbol{k}} = \sqrt{\frac{|\omega_k|}{2}}\, \chi_{\boldsymbol{k}} + \frac{i}{\sqrt{2\,|\omega_k|}}\, p_{\boldsymbol{k}}\, ,
\\
 &a^{\dagger}_{-\boldsymbol{k}} = \sqrt{\frac{|\omega_k|}{2}}\, \chi_{\boldsymbol{k}} - \frac{i}{\sqrt{2\,|\omega_k|}}\, p_{\boldsymbol{k}}\, .
\end{split}
\end{equation}
Substituting the expressions (\ref{umass}) and (\ref{pmass}), and comparing with (\ref{bogo}) one gets the following expressions for $\alpha_k$ and $\beta_k$ in terms of the mode functions $u_k$:
\begin{equation}
\label{alpha}
    \alpha_k = \sqrt{\frac{|\omega_k|}{2}}\,u_k + \frac{i}{\sqrt{2\,|\omega_k|}}\,u^\prime_k\,,
\end{equation}
\begin{equation}
\label{beta}
    \beta_k = \sqrt{\frac{|\omega_k|}{2}}\,u^*_k + \frac{i}{\sqrt{2\,|\omega_k|}}\,u^{*\,\prime}_k\,.
\end{equation}
With these definitions for the Bogoliubov coefficients, and using (\ref{normalization}), we can compute a quantity which will be useful later:
\begin{equation}
\label{beta2mass}
    |\beta_k|^2 = \frac{|\omega_k|}{2}\, |u_k|^2 + \frac{1}{2\,|\omega_k|}\,|u_k^\prime|^2 - \frac{1}{2}\,.
\end{equation}

Having introduced the Bogoliubov coefficients, we need to interpret these quantities in terms of physical observables. Their evolution can be understood in terms of particle creation by a spacetime-dependent background \cite{birrell_davies_1982, Mukhanov:2007zz, Agullo:2022ttg, Lueders:1990np}. 
Consider two sets of ladder operators ($a$,$a^\dagger$) and ($b$,$b^\dagger$) and their corresponding vacua defined through $a|0\rangle_a =0$ and $b|0\rangle_b =0$. These sets can be linked via a constant Bogoliubov transformation:
\begin{equation}
\begin{split}
 &b = A\,a + B\,a^\dagger\,,
\\
 &b^\dagger = A^*\,a^\dagger + B^*\,a\,.
\end{split}
\end{equation}
The two vacua $|0\rangle_a $ and $|0\rangle_b $ do not necessarily coincide. The expectation value of the number operator of one basis with respect to the vacuum of the other basis is given by $|B|^2$. In other words, an observer in the $a$-basis feels the vacuum as filled with $b$-particles.

In the cosmological case, where time-dependent Bogoliubov transformations has been introduced, the instantaneous vacuum defined by the time-dependent ladder operators ($a_{\boldsymbol{k}}(\tau)$, $a_{\boldsymbol{k}}^\dagger(\tau)$) is filled with particles associated with the initial-time operators ($a_{\boldsymbol{k}}^0$, $a_{\boldsymbol{k}}^{\dagger\,0}$). 
However, unlike in Minkowski space, 
in a general curved spacetime there is no way to uniquely define a vacuum state. This ambiguity can be overcome if the effective spacetime seen by the fluctuations approaches Minkowski both in the far past and in the far future, namely, if adiabatic vacua exist asymptotically such that the mode functions are given by plane waves.
There the fluctuations can be interpreted as particles, and the Bogoliubov coefficients relating the ladder operators in the asymptotic past and future give the final particle number as~$|\beta_k|^2$.

To verify the above discussion explicitly, let us compute the expectation value of the Hamiltonian~(\ref{Hmass}). Using (\ref{u}) and (\ref{p}), we can rewrite the Hamiltonian in terms of the ladder operators as
\begin{equation}
\label{Hbeta}
\begin{split}
    \mathcal{H} = \int \frac{d^3k}{2}\,&\left[ \left( \frac{|\omega_k|}{2} + \frac{\omega_k^2}{2\,|\omega_k|} \right) \left(2\, a^\dagger_{\boldsymbol{k}} a_{\boldsymbol{k}} + \left[a_{\boldsymbol{k}}, a^\dagger_{\boldsymbol{k}}\right] \right)  \right. \\
    & \left.~~ - \left( \frac{|\omega_k|}{2} - \frac{\omega_k^2}{2\,|\omega_k|} \right) \left( a^\dagger_{\boldsymbol{k}} a^\dagger_{-\boldsymbol{k}} + a_{\boldsymbol{k}} a_{-\boldsymbol{k}} \right) \right] .
\end{split}
\end{equation}
Since we are interested in the late time limit of (\ref{Hbeta}), we can simply drop the absolute value for $\omega_k$ and simplify the previous expression. In fact during radiation domination the scale factor follows $a''/a =0$, and after the onset of the oscillations the effective mass squared is positive, hence $\omega_k^2$ becomes positive.
Taking the expectation value over the initial vacuum, i.e. the vacuum annihilated by the initial time-independent annihilation operator, $a_{\bd{k}}^0 |0 \rangle  = 0$ for all $\boldsymbol{k}$, we get:
\begin{equation}
\label{meanH}
    \langle\mathcal{H}\rangle = \int d^3k\, \omega_k\, \langle a^\dagger_{\boldsymbol{k}} a_{\boldsymbol{k}} \rangle + \text{(vacuum energy)}.
\end{equation}
The divergent vacuum energy can be removed by a normal ordering. The expectation value for the ladder operators can be computed from (\ref{bogo}) as follows:
\begin{equation}
    \langle a^\dagger_{\boldsymbol{k}} a_{\boldsymbol{k}} \rangle = |\beta_k|^2 \langle a^0_{-\boldsymbol{k}}\, a^{0\,\dagger}_{-\boldsymbol{k}} \rangle = |\beta_k|^2 \frac{V}{(2 \pi)^3}\,,
\label{eq:6.18}
\end{equation}
where $V = \int d^3x = (2 \pi)^3\,\delta^3(0)$ is the comoving volume.
The expression for $|\beta_k|^2$ is given in terms of the mode functions, as in (\ref{beta2mass}). Notice that the particle interpretation is valid only when the perturbation mode does not feel the expansion, i.e. in the sub-horizon limit. In this limit, $\omega_k$ is positive and can be interpreted as the frequency of the perturbation mode. $|\beta_k|^2$ can then be interpreted as the expectation value of the number of particles with comoving momentum~$k$, created per comoving volume due to the time-dependent background. The divergent factor $\delta^3(0)$ arises because we are considering an infinite spatial volume, but the mean density of particles per comoving volume is finite. We therefore expect the norm of $\beta_k$ to approach a constant in the asymptotic future.

\subsection{Squeezing Parameters}
\label{squeezingparameters}
Thanks to the constraint (\ref{commu1}), the two complex Bogoliubov coefficients can be parameterised by three real variables, the so-called squeezing parameters, as:
\begin{equation}
\label{sqpar}
\begin{split}
 &\alpha_{k}(\tau) = e^{-i\vartheta_{k}(\tau)}\, \cosh{r_{k}(\tau)}\, ,
\\
 &\beta_{k}(\tau) = e^{i\left[ \vartheta_{k}(\tau) + 2\varphi_{k}(\tau) \right]}\, \sinh{r_{k}(\tau)}\, ,
\end{split}
\end{equation}
where $r_k$ is restricted to be non-negative.
Inverting these relations yields the expressions for the squeezing parameters in terms of the Bogoliubov coefficients:
\begin{equation}
\label{sqparf}
\begin{split}
 &r_k = \text{arcsinh} \abs{\beta_k}\,,
 \\
 &\vartheta_k = -\arg \left(\alpha_k \right)\,,
 \\
 & \varphi_k  = \frac{1}{2} \arg \left( \alpha_k \beta_k \right)\,.
\end{split}
\end{equation}

The process of particle creation explained in Section \ref{bogotransformation} can be equivalently described by means of the squeezing formalism. In particular, the cosmological particle creation amounts to squeezing the vacuum. The physical meaning of the squeezing parameters can be understood studying a quadrature pair in phase space of the physical system of interest. A squeezed state is given by a quadrature pair for which the uncertainty in one variable is small, while the uncertainty in the other variable is correspondingly large, in order to respect the Heisenberg uncertainty principle. This state is described by a rotated ellipse in phase space. The $r_k$~parameter is directly linked with the semi-axes of this ellipse, while the $\varphi_k$~parameter is related to its orientation. In particular, the $r_k$~parameter tells us how much the ellipse is squeezed along one direction while elongated along the other. The $\vartheta_k$~angle is actually unphysical. A detailed description of the physical meaning of the squeezing parameters is given in Appendix \ref{sqparphysmean}.

The squeezing parameters are also linked to cosmological observables. We have seen in Section \ref{bogotransformation} that $|\beta_k|^2$ can be interpreted as the number of particles created by the expansion of spacetime. The parameter $r_k$ is given in terms of this net number of particles, as shown in (\ref{sqparf}). The parameter $\varphi_k$ is instead linked to the power spectrum of the field fluctuations. As explained in Appendix \ref{sqparphysmean}, the initial state we are working with, i.e. the vacuum state, is rotationally invariant; hence only one phase is physically meaningful. 
From the expectation value of the field fluctuation:
\begin{equation}
\label{PSdeltaphi}
 \langle \delta \phi^2 \rangle
= \int \frac{d^3 k}{(2 \pi)^3} \frac{\abs{u_k}^2}{a^2} = \int \frac{dk}{k}\, P_k\,,
\end{equation}
the power spectrum is written as,
\begin{equation}
\label{power}
    P_k= \frac{k^3}{2\pi^2\,a^2}\,\abs{u_k}^2\,.
\end{equation}
Substituting the expression for $|u_k|^2$ in terms of the Bogoliubov coefficients (\ref{alpha}) and (\ref{beta}), and exploiting the constraint (\ref{commu1}), the power spectrum can be rewritten in terms of $|\beta_k|$ and $\arg \left(\alpha_k\beta_k \right)$:
\begin{equation}
\label{powersq}
\begin{split}
    P_k &= \frac{k^3}{2\pi^2\,a^2}\,\frac{\abs{\alpha_k +\beta_k^*}^2}{2\abs{\omega_k}} \\
    &= \frac{k^3}{4\pi^2\,a^2\,\abs{\omega_k}}\, \left[1 + 2\abs{\beta_k}^2 + 2\abs{\beta_k}\, \sqrt{1+\abs{\beta_k}^2}\, \cos \left( \ro{arg}\left( \alpha_k\beta_k \right) \right) \right]\,.
\end{split}
\end{equation}
In terms of the squeezing parameters, $\arg \left(\alpha_k\beta_k \right)$ corresponds to $\varphi_k$, which is therefore a physical quantity.

\section{Background Evolution}
\label{background}

To study the evolution in time of the Bogoliubov coefficients and then in turn of the squeezing parameters, we first analyse the background field dynamics by numerically solving the homogeneous equation of motion~(\ref{phi0}). It is convenient to introduce the background displacement angle 
\begin{equation}
\theta = \bar \phi/f,
\end{equation}
and solve the equation in terms of this variable. In this way we do not need to specify the value of the symmetry breaking scale $f$. The equation of motion for the background angle in conformal time is given by:
\begin{equation}
    \label{backtheta}
    \theta^{\prime\prime} + 2aH\theta^\prime + a^2m^2 \sin \theta = 0\,.
\end{equation}

\begin{figure}[t]
    \centering
    \begin{subfigure}[t]{0.477\textwidth}
        \includegraphics[width=\textwidth]{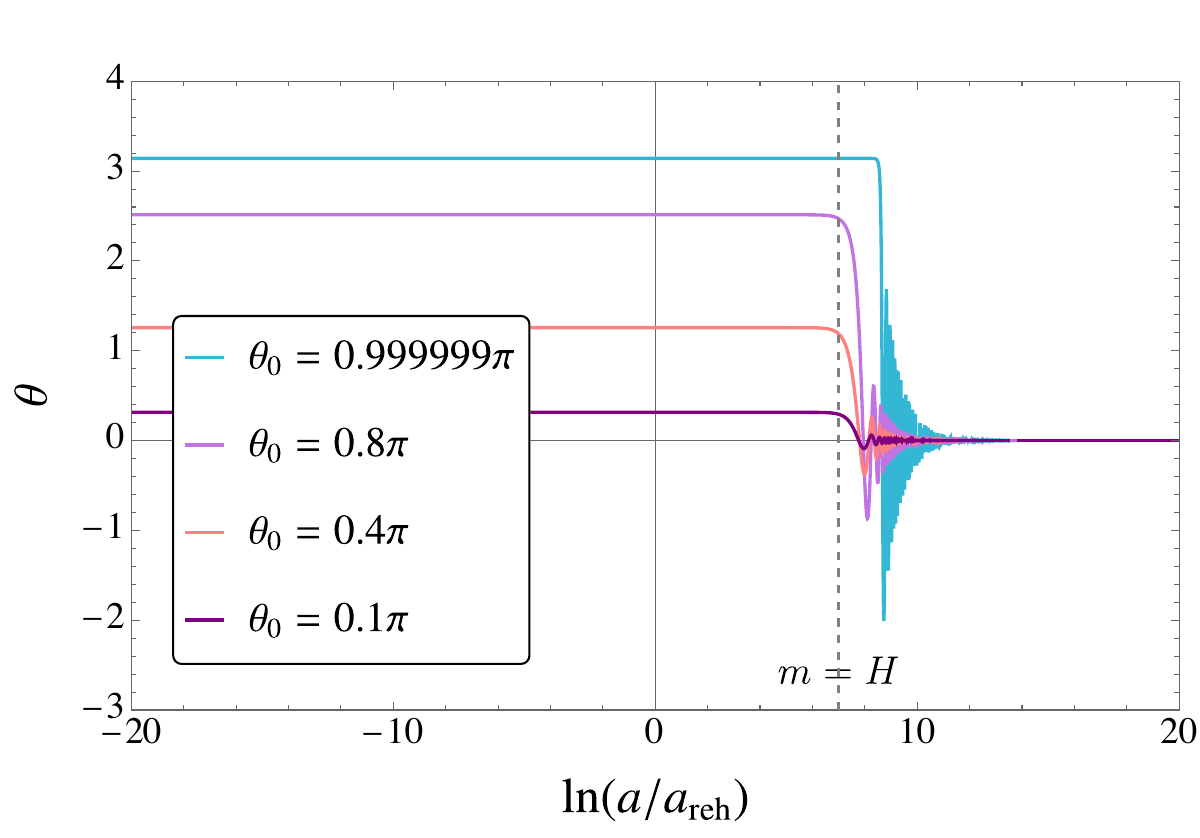}
        \label{backcondini}
    \end{subfigure}
    \quad
    \begin{subfigure}[t]{0.483\textwidth}
        \includegraphics[width=\textwidth]{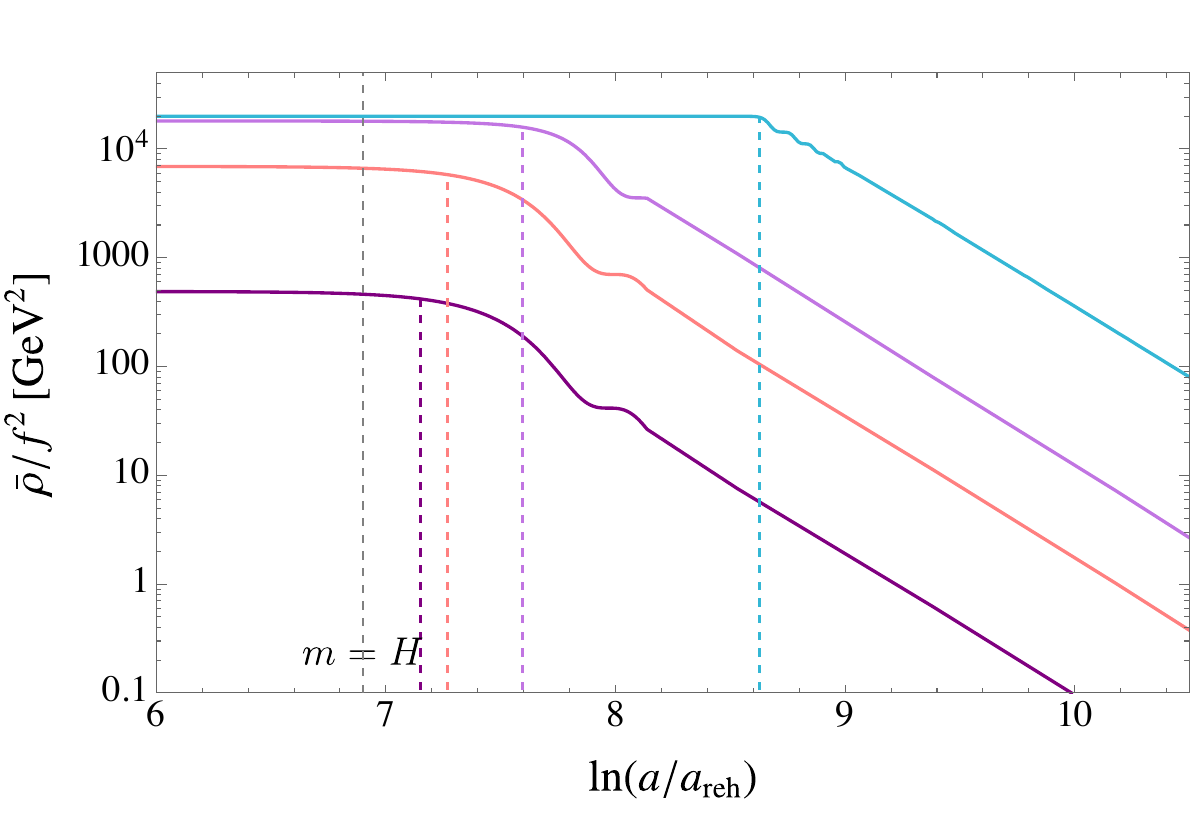}
        \label{ed}
    \end{subfigure}
    \caption{Time evolution of the background field (left) and of the background energy density (right) for different initial field values. The values of the initial angle are specified in the legend in the left plot. The parameters chosen are $H_{\ro{inf}}=10^8\,\text{GeV}$ and $m=10^2\,\text{GeV}$. The grey dashed vertical line shows the time when the axion mass becomes equal to the Hubble rate. Dashed vertical lines in other colours denote the onset of the axion oscillation for the different initial displacement angles.}
    \label{backgroundtimeevolution}
\end{figure}

We expect the axion background to start oscillating after the Hubble rate becomes comparable with the axion mass.
Solving the equations of motion numerically for different initial field values during inflation, we obtain the results reported in Figure \ref{backgroundtimeevolution} (left). Here we considered the Hubble rate in equation (\ref{Hubbletau}) with $H_{\ro{inf}}=10^8\, \text{GeV}$, and took the mass as $m=10^2\, \text{GeV}$. Then the third term in (\ref{backtheta}) becomes relevant at around $\ln (a / a_{\ro{reh}}) = \ln (H_{\ro{inf}} / m)^{1/2} \approx 7$, as shown in Figure \ref{backgroundtimeevolution} (left). The different colours represent different initial angles as shown in the legend, while the vertical dashed line identifies when the mass of the axion field becomes comparable with the Hubble rate. If we choose as the initial field value $\theta_0 =0.1\, \pi$, it is close enough to the minimum 
such that the axion potential is well approximated by a quadratic.
Indeed, if we compute the background trajectory with 
a quadratic potential  we find the same result. Looking at Figure \ref{backgroundtimeevolution} (left) we see that starting close to the minimum, the amplitude of the oscillations is reduced. A huge difference in both the amplitude of the oscillations and the time of the onset of the oscillations can be seen if we set the initial field value to be very close to the hilltop.

\begin{figure}[!htbp]
    \centering
    \includegraphics[scale=.6]{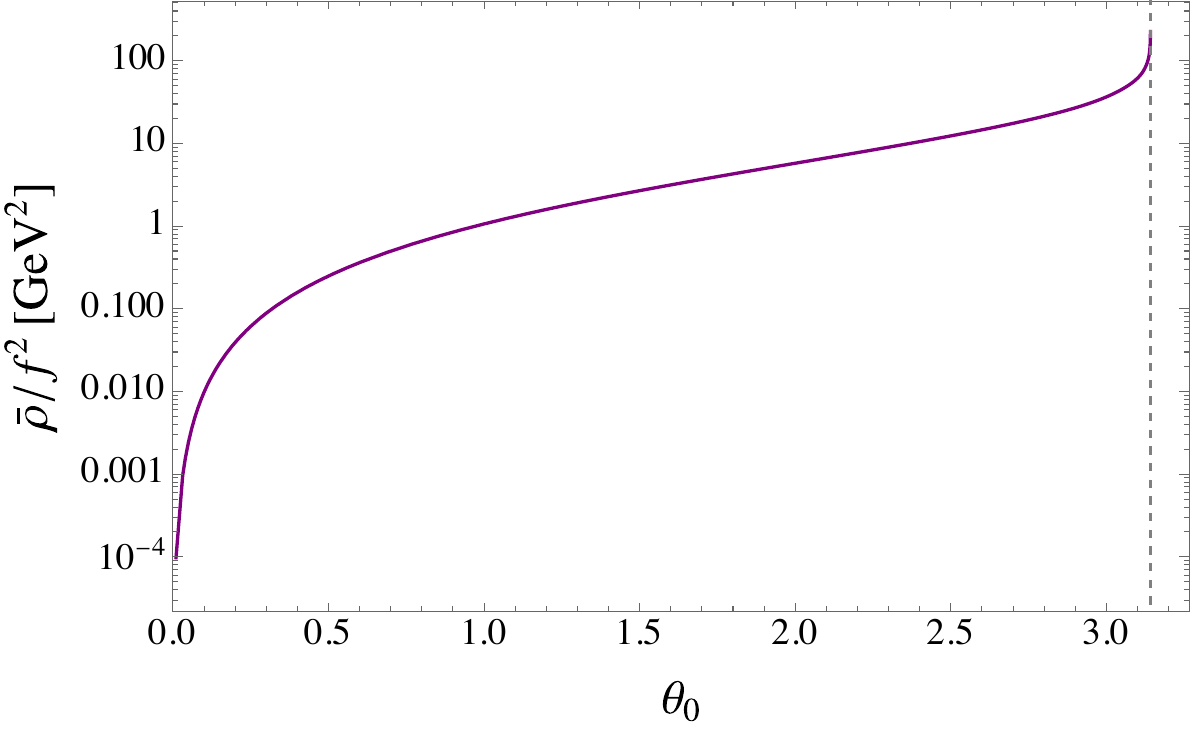}
    \caption{Asymptotic value of the energy density of the background field $\bar\rho$ as function of the initial displacement angle $\theta_0$. The parameters used are $H_\ro{inf} = 10^8\,\text{GeV}$ and $m=10^2\,\text{GeV}$.}
    \label{endensfinal}
\end{figure}

This behaviour can be further understood by looking at the energy density of the background field. The energy density can be obtained as
\begin{equation}
\label{rhoback}
    \bar\rho = \frac{\dot {\bar\phi}^2}{2} + V(\bar\phi) = \frac{f^2}{2a^2}\, \theta^{\prime\,2} + V(\theta f).
\end{equation}
Given the solution for the background field evolution, the energy density can be directly computed. The time evolution of the energy density normalized by $f^2$ is reported in Figure~\ref{backgroundtimeevolution} (right), which zooms into the RD epoch around the time when the oscillations begin.
The energy density is constant until the background field starts oscillating; thereafter it decays as $a^{-3}$. 
The vertical dashed lines in Figure \ref{backgroundtimeevolution} (right) show when the oscillations begin; this is obtained as the time when the energy densities extrapolated from the asymptotic behaviours in the past ($\bar{\rho} = \ro{const.}$) and in the future ($\bar{\rho} \propto a^{-3}$) cross each other.
The value of the energy density in the asymptotic future in terms of the initial displacement angle is reported in Figure \ref{endensfinal}. It has be computed at the final time $\ln(a/a_\ro{reh})=10$. The final numerical value of $\bar\rho/f^2$ itself is not relevant for our discussion. What is relevant is that the energy density becomes strongly enhanced towards the  hilltop. This behavior can be understood by noticing that the field starts oscillating later as one approaches the hilltop.


\section{Evolution of Axion Fluctuations}
\label{axionfluct}

\subsection{Mode Functions}
\label{modefunctions}
In order to evaluate the evolution in time for the Bogoliubov coefficients and the squeezing parameters, we need firstly to solve the equation of motion for the mode functions, given by:
\begin{equation}
\label{modeeq}
    u_k'' + \left( k^2 + m_{\ro{eff}}^2 a^2 - \frac{a''}{a} \right) u_k = 0\,.
\end{equation}
If the potential has only the mass term, an analytical solution can be obtained, while if we work with the full cosine potential we have to rely on numerical computations. The analytical derivation for a mass term potential is reported in Appendix \ref{quadraticmodes}.

The time evolution of the amplitude of the mode functions is reported in Figure~\ref{modes0109}. It has been numerically computed for two different values of the initial background field during inflation: 
$\theta_0 =0.1\,\pi$ and $\theta_0 =0.999999\,\pi$. 
The initial condition for the mode function is set such that it starts as a positive-frequency WKB solution obeying the normalization condition~(\ref{normalization}), when the wave mode is deep inside the horizon during inflation.
The parameters chosen for the evaluation are $H_\ro{inf}=10^8\, \text{GeV}$, $m=10^2\, \text{GeV}$ and $k/a_\ro{reh}=10^2\,\text{GeV}$. These values have been chosen in order to consider a mode that leaves the horizon during inflation, and stays outside the horizon when the background field starts oscillating around the minimum, during radiation domination. The dashed vertical lines in Figure \ref{modes0109} indicate the time of the horizon exit, reheating and the onset of the oscillations of the background field.

Looking at Figure \ref{modes0109}, we see that the mode function's amplitude stays constant until horizon exit, then grows outside the horizon until the background field starts to oscillate. This behaviour is more or less independent of the initial misalignment angle (and thus the lines overlap). Instead, at the onset of the oscillations the amplitude of the mode functions differ based on the value of the initial displacement angle. If the initial field is close to the minimum, then the approximation to the quadratic potential is enough to describe the system. The time evolution obtained from the analytical solution for a quadratic potential matches the time evolution for an initial field value $\theta_0 = 0.1 \pi$. However, if the initial field is close to the hilltop an extra enhancement in the amplitude of the mode functions is observed due to the anharmonic effects.

\subsection{Bogoliubov Coefficients}
\label{bogoevolution}

\begin{figure}[t]
    \centering
    \includegraphics[scale=.6]{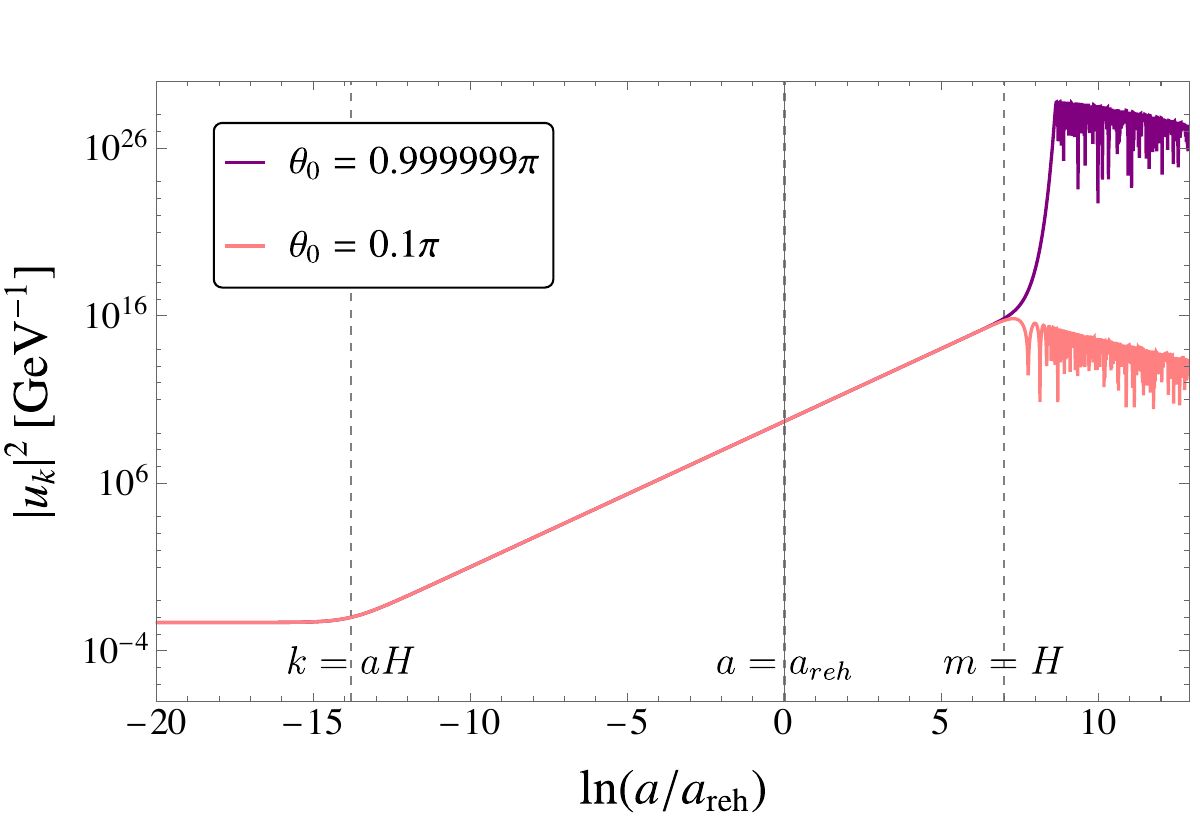}
    \caption{Evolution in time of the amplitude of the mode functions for a wave mode $k/a_\ro{reh}=10^2\, \text{GeV}$, and for two different initial values of the background field. An increase in $|u_k|^2$ is observed approaching the hilltop of the potential. Dashed vertical lines denote when the mode exits the horizon, reheating and the onset of the oscillations, respectively. The values chosen for the Hubble rate and the axion mass are $H_\ro{inf}=10^8\, \text{GeV}$ and $m=10^2\,\text{GeV}$.}
    \label{modes0109}
\end{figure}

\begin{figure}[!ht]
    \centering
    \includegraphics[scale=.6]{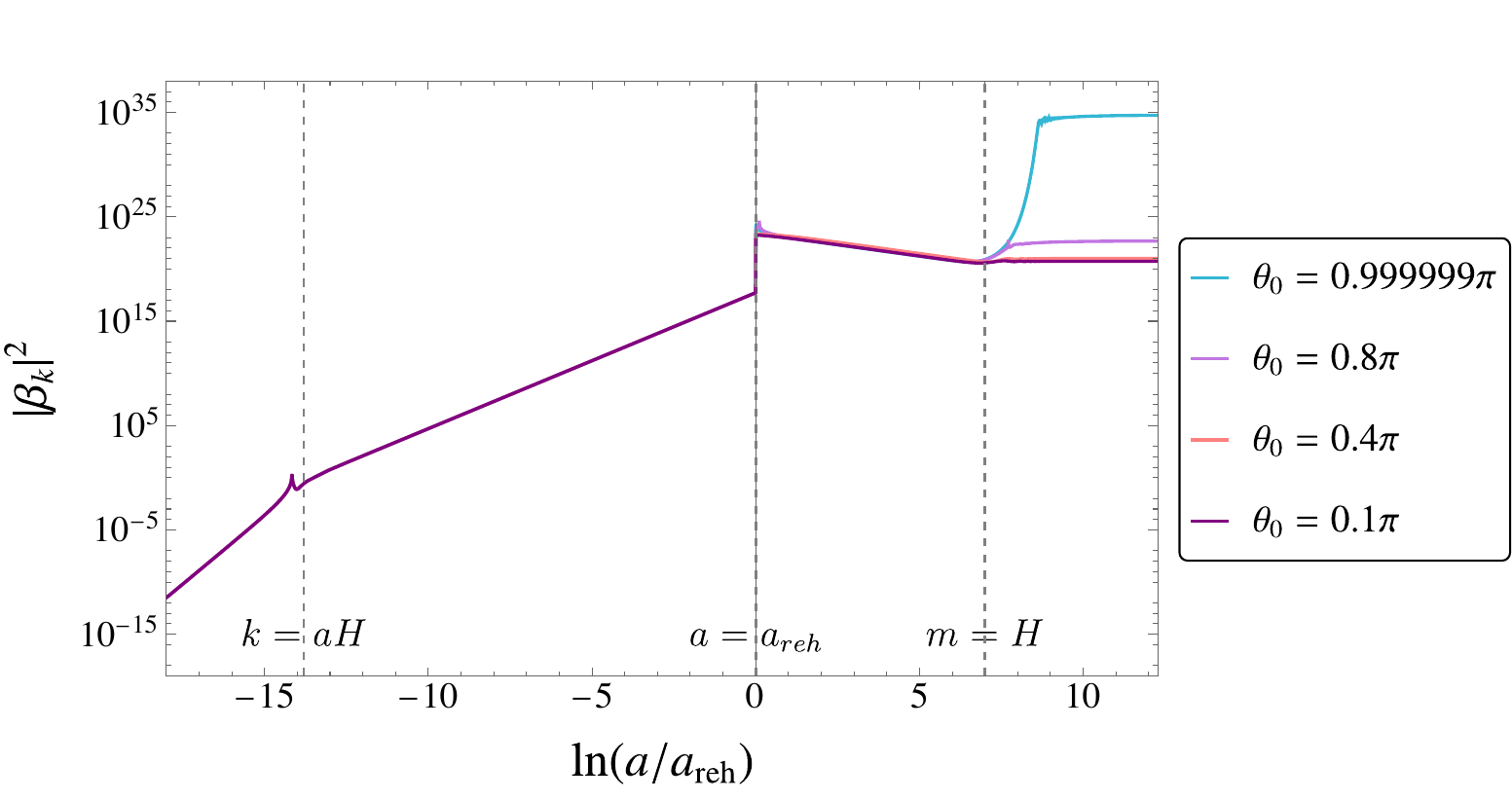}
    \caption{Evolution in time of the modulus square of the $\beta_k$-coefficient for few different initial values of the background field. The parameters chosen are the same as Figure \ref{modes0109}. When $\theta_0$ approaches the hilltop of the potential the final value of $|\beta_k|^2$ significantly increases. Dashed vertical lines denote when the mode exits the horizon, reheating and the onset of the oscillations, respectively.}
    \label{betacosine}
\end{figure}

Exploiting relations (\ref{alpha}), (\ref{beta}) and (\ref{sqparf}), it is possible to evaluate the evolution in time of the Bogoliubov coefficients and in turn of the squeezing parameters. We start our analysis from the Bogoliubov coefficients, and specifically from $|\beta_k|^2$, which is directly linked with physical quantities, as explained in Section \ref{bogotransformation}.
The numerical results are reported in Figure \ref{betacosine}, for the same values of parameters used in Figure~\ref{modes0109} for the amplitude of the mode functions.
The modulus square of the beta coefficient approaches zero in the asymptotic past, when the mode is far inside the horizon. Moreover it approaches a constant value in the asymptotic future, and specifically it becomes constant when the axion field starts oscillating.
We further notice that the behaviour of $|\beta_k|^2$ changes depending on the initial field value. The evolution in time is more or less independent of the initial condition while the axion field is frozen by the Hubble friction; however, when the potential term becomes comparable with the Hubble rate, the field dynamics right before the onset of the oscillations affects the asymptotic value of the Bogoliubov coefficients.
We see in Figure \ref{betacosine} that, approaching the hilltop of the potential, the asymptotic value of $|\beta_k|^2$ increases. 
On the other hand with $\theta_0=0.1\,\pi$, the evolution of $|\beta_k|^2$ follows that for a quadratic potential, which is also analytically studied in Appendix~\ref{quadraticmodes}. 


The spikes in the evolution of $|\beta_k|^2$ seen in the plots, as well as the discontinuous jump at reheating, can be understood in terms of the frequency~(\ref{omega}). In the asymptotic past, $\omega^2_k$ is dominated by $k^2$. When the perturbation leaves the horizon during inflation, at $k=aH_{\ro{inf}}$, the frequency squared reduces to $\omega_k^2 \simeq -k^2$. This implies that right before the horizon exit the frequency squared has turned from positive to negative. This gives a divergence in $|\beta_k|^2$, since it is defined in (\ref{beta2mass}) with $\omega_k$ in the denominator. The spike after reheating has a similar explanation. During RD, $\omega_k^2 = k^2 + m^2_\ro{eff}a^2$; the $k^2$ term is always positive, while $m^2_\ro{eff}$ could be either positive or negative, depending on the value of the cosine. If the initial field is close to the hilltop, then $m^2_\ro{eff}$ is originally negative. With the parameter choice in the plot ($m=k/a_\ro{reh}=10^2\,\text{GeV}$), the tachyonic mass term dominates the frequency right after reheating. This implies that the frequency squared flips its sign immediately after reheating, causing the spike in $|\beta_k|^2$. 
There is another spike at the onset of the oscillations, depending on the initial displacement angle. In fact, when the background field starts oscillating,  $m^2_\ro{eff}$ goes from negative to positive. Since $\omega^2_k \simeq m^2_\ro{eff} a^2$ at that time, it crosses zero, making $|\beta_k|^2$ diverge.
The jump at reheating is instead caused by the discontinuity of the frequency $\omega_k$, which arises due to $a''/a$ suddenly vanishing upon instantaneous reheating. 

However we remark that the spikes and jumps of $|\beta_k|^2$ appear only while adiabaticity is violated and hence $|\beta_k|^2$ cannot be interpreted as the particle number, as we will explain in the next subsection. The irregular behaviours are artifacts due to our definition of~$|\beta_k|^2$, as well as the choice of the cosmological background, and hence should not have physical consequences. We actually show in Appendix~\ref{morerealistic} that a non-instantaneous reheating removes the jump of~$|\beta_k|^2$, but does not alter the final value of~$|\beta_k|^2$.

\subsubsection{Adiabaticity Condition}
\label{adiabaticitycondition}
We have seen that $|\beta_k|^2$ approaches a constant when the background field starts oscillating.
In order to understand why this is the case, let us analyse the so-called adiabaticity condition, which is given by:
\begin{equation}
\label{adiacond}
    \bigg|\, \frac{\omega'_k}{\omega_k^2}\, \bigg|^2\,, \bigg|\, \frac{\omega''_k}{\omega_k^3}\, \bigg| \ll 1\,.
\end{equation}
Here $\omega_k$ is the effective frequency associated with the mode, defined in (\ref{omega}).
If $\omega_k$ is real and the adiabaticity condition holds, then the solution to the equation of motion for the perturbations should approach a plane wave solution. In particular, it implies that the WKB approximation holds so that the mode function can be written as
\begin{equation}
\label{WKB}
    u_k(\tau) \simeq \frac{A_k}{\sqrt{2\omega_k(\tau)}}\, e^{-i\,\int^\tau \omega_k(\tau') d\tau'} + \frac{B_k}{\sqrt{2\omega_k(\tau)}}\, e^{i\,\int^\tau \omega_k(\tau') d\tau'}\,.
\end{equation}
When this is the case, the time-dependent Bogoliubov coefficients $\alpha_k$ and $\beta_k$ reduce to the C-numbers $A_k$ and $B_k$, up to a time-dependent phase.

Let us analytically assess the adiabaticity condition for a quadratic potential, i.e. $m_{\ro{eff}}^2 = m^2$.
In the dS epoch, the first adiabaticity parameter $|\omega'_k/\omega^2_k|$ is tiny before the mode exits the horizon, however when the mode is far outside the horizon then it settles to a constant value of $1/\sqrt{2}$. This can be shown by ignoring the axion mass which is much smaller than the inflationary Hubble scale, giving
\begin{equation}
\label{adiacondDS}
    \left| \frac{\omega_k'}{\omega_k^2} \right| \simeq 
\frac{1}{\sqrt{2}} 
\left| 1 - \frac{k^2}{2 a^2 H_{\ro{inf}}^2} \right|^{-3/2}\,.
\end{equation}
The second adiabaticity parameter $\omega_k'' / \omega_k^3$ exhibits a similar behaviour.

The adiabaticity condition becomes satisfied again after inflation when the axion field starts oscillating. During radiation domination, the first adiabaticity parameter reduces to
\begin{equation}
\label{domegaRD}
    \frac{\omega_k'}{\omega_k^2} = \frac{a^3 H\, m^2}{\left( k^2 + a^2m^2 \right)^{3/2}}.
\end{equation}
Given this result, there are two possible regimes:
\begin{equation}
\label{adiacondRD}
\frac{\omega_k'}{\omega_k^2} \, \simeq
\begin{cases}
    \frac{a^3H\,m^2}{k^3} & \quad k \gg a\,m, \\
    \frac{H}{m} & \quad k \ll a\,m.
\end{cases}
\end{equation}
We restrict ourselves to wave modes that are outside the horizon at the onset of the oscillations, i.e. $k < a_{\ro{osc}} m$. 
Hence after the oscillation begins, namely when $a > a_{\ro{osc}}$ and $H < m$, the adiabaticity parameter is given by the second line, which is smaller than unity.
Therefore starting from that point the adiabaticity condition is always recovered. The behaviour long before $a_\ro{osc}$ depends on the actual value of $k$. The second adiabaticity parameter is
\begin{equation}
    \frac{\omega_k''}{\omega_k^3} = \frac{m^2}{\left( k^2 + a^2m^2 \right)^{2}}\, \left( H^2a^4 - \frac{m^2H^2a^6}{k^2 + a^2m^2} \right).
\end{equation}
Considering again the two possible regimes, we have:
\begin{equation}
    \frac{\omega_k''}{\omega_k^3} \, \simeq
\begin{cases}
    \cfrac{m^2a^4H^2}{k^4} & \quad k \gg a\,m, \\
    \cfrac{k^2 H^2}{a^2 m^4} & \quad k \ll a\,m.
\end{cases}
\end{equation}
Also in this case the condition is surely satisfied when $a>a_\ro{osc}$.

The above discussions apply similarly to the full cosine potential, for which we numerically checked with initial misalignment angles close to the hilltop. The behaviours of the adiabaticity parameters are the same while the axion is frozen by the Hubble friction.
Soon after the axion starts to oscillate the adiabaticity parameters also oscillate, however, within a few $e$-folds the adiabaticity condition becomes satisfied.

\subsubsection{Enhancement in the Number of Particles}
\label{enhancement}

\begin{figure}[t]
    \centering
    \begin{subfigure}[t]{0.48\textwidth}
        \includegraphics[width=\textwidth]{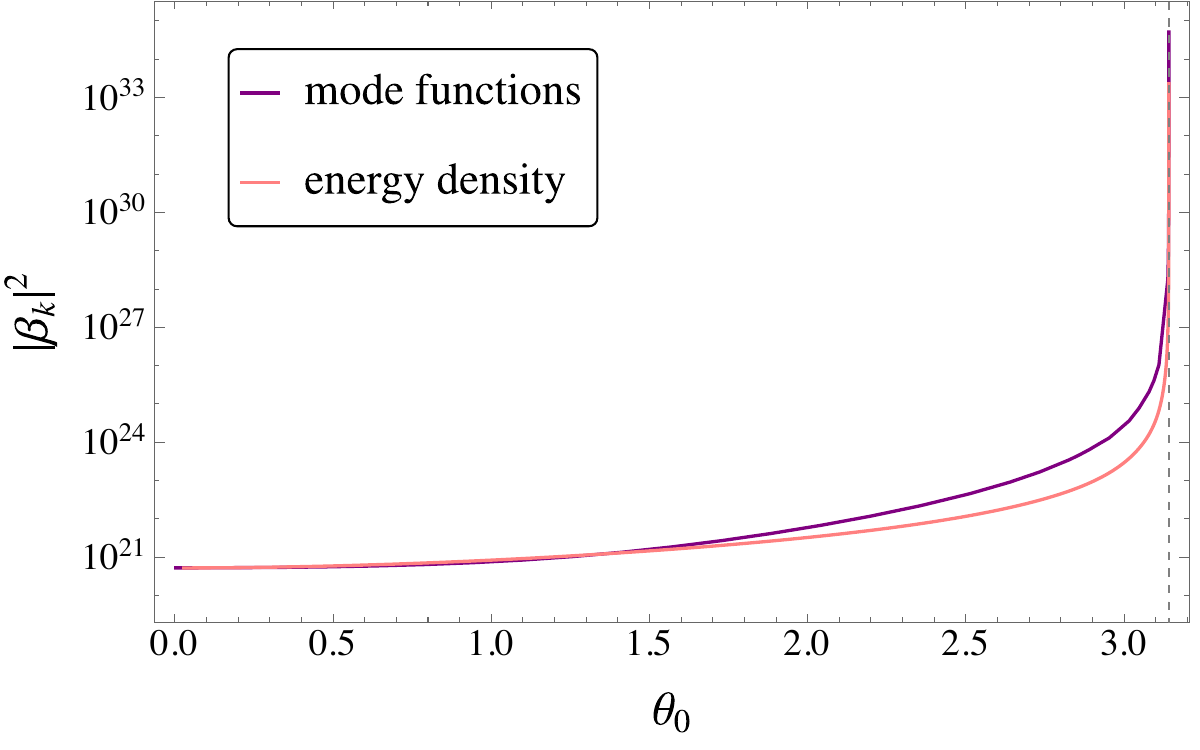}
        \label{betacondini}
    \end{subfigure}
    \quad
    \begin{subfigure}[t]{0.48\textwidth}
        \includegraphics[width=\textwidth]{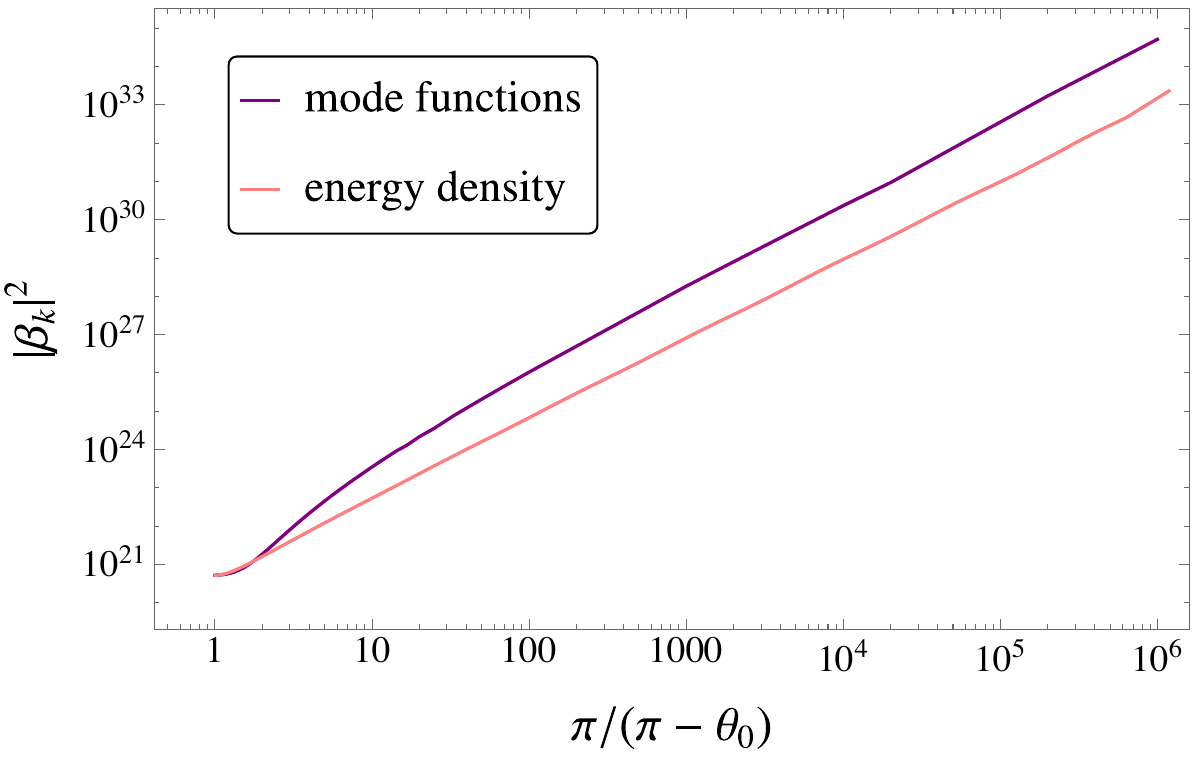}
        \label{betaxfinalpoints}
    \end{subfigure}
    \caption{Asymptotic value of $|\beta_k|^2$ as functions of the initial displacement angle (left), and of $\pi/(\pi-\theta)$ (right). The parameters are chosen as $H_{\ro{inf}}=10^8\, \ro{GeV}$, $m=10^2\, \ro{GeV}$, and $k/a_\ro{reh}=10^2\,\text{GeV}$. The purple line refers to $|\beta_k|^2$ computed through the mode functions. The pink line shows the second derivative of the axion density in terms of the initial field value, as given in the right-hand side of~(\ref{eq:star}).}
    \label{betacondiniall}
\end{figure}

In order to understand the enhancement in the beta coefficient due to the anharmonic effects, we show in Figure \ref{betacondiniall} (left, purple line) the value of $|\beta|^2$ in the asymptotic future as a function of the initial field value $\theta_0 = \bar \phi_0/f$.
(The pink line will be explained in Section \ref{betaEnDens}.)
The asymptotic value of the Bogoliubov coefficient significantly increases as one approaches the hilltop at $\theta_0 = \pi$. This behaviour can be interpreted analysing the time at which the field starts oscillating. As we have seen in Section \ref{background}, increasing the initial field value delays the onset of the oscillations. Approaching the hilltop this delay becomes more and more evident. If we consider the field perturbation around a background value close to the minimum, the delay in the onset of the oscillation is too tiny to affect the evolution in time of $\delta \phi_k$.
However, if the field perturbation is considered around a background field value close to the maximum of the potential, then the evolution in time of the perturbation changes dramatically.
In fact, close to the hilltop even a small difference in the initial position leads to a huge difference in the time of the onset of the oscillations. This implies that patches of the Universe that differ by a tiny variation of the initial misalignment angle will start to oscillate at very different times, sourcing huge fluctuations.
This in turn reflects into the modulus square of the mode functions, that enters directly into $|\beta_k|^2$. In the hypothetical situation where the angle is set initially to~$\pi$, the field does not start oscillating at all. Hence $\delta\phi_k$ is going to become infinite, thus making $|\beta_k|^2$ divergent. 

To better analyse the behaviour of $|\beta_k|^2$ near the hilltop it is worth working in terms of $\pi/(\pi-\theta_0)$. In terms of this variable, the hilltop at $\theta_0 = \pi$ is shifted to $\infty$. In Figure~\ref{betacondiniall} (right, purple line), $|\beta_k|^2$ is reported in terms of this new variable. 
We find that the asymptotic value of $\abs{\beta_k}^2$ grows towards the hilltop\footnote{A similar scaling was also found for the two-point correlator of the field fluctuations in the context of axionic curvatons~\cite{Kawasaki:2011pd}.}
as $(\pi - \theta_0)^{-2}$.

We should remark that the above discussions do not apply arbitrarily close to the hilltop. In particular if the initial displacement from the hilltop, $\pi - \theta_0$, is comparable to or smaller than the angular quantum fluctuation $\delta \theta$ (whose amplitude is of $  H_{\ro{inf}} / f$), 
then the perturbative expansion around the classical background breaks down.
In this regime the field can roll down to the other side of the hill 
in some patches of the universe, and give rise to axionic domain walls. In this work we do not consider such cases.

\subsubsection{Parameter Dependence}
\label{enhancement}

Let us now study the dependence of $|\beta_k|^2$ on the parameters $H_{\ro{inf}}$, $m$, $k$. The dependence on the parameters can be derived analytically for a quadratic potential. The complete derivation is reported in Appendix~\ref{quadraticexpansion}, where we show that the asymptotic value of $|\beta_k|^2$ for a quadratic potential is given by
\begin{equation}
\label{betaexpminimum}
    |\beta_k|^2 = \frac{1}{8\pi}\, \Gamma(1/4)^2\, \left(\frac{a_\ro{reh}}{k} \right)^3 \frac{H_{\ro{inf}}^{7/2}}{\sqrt{m}}\,.
\end{equation}
A physical interpretation of this result in terms of the energy density will be presented in Section \ref{betaEnDens}. The dependence of $|\beta_k|^2 \propto k^{-3}$ shows the scale invariance of the perturbation.

\begin{figure}[t]
    \centering
    \includegraphics[scale=.5]{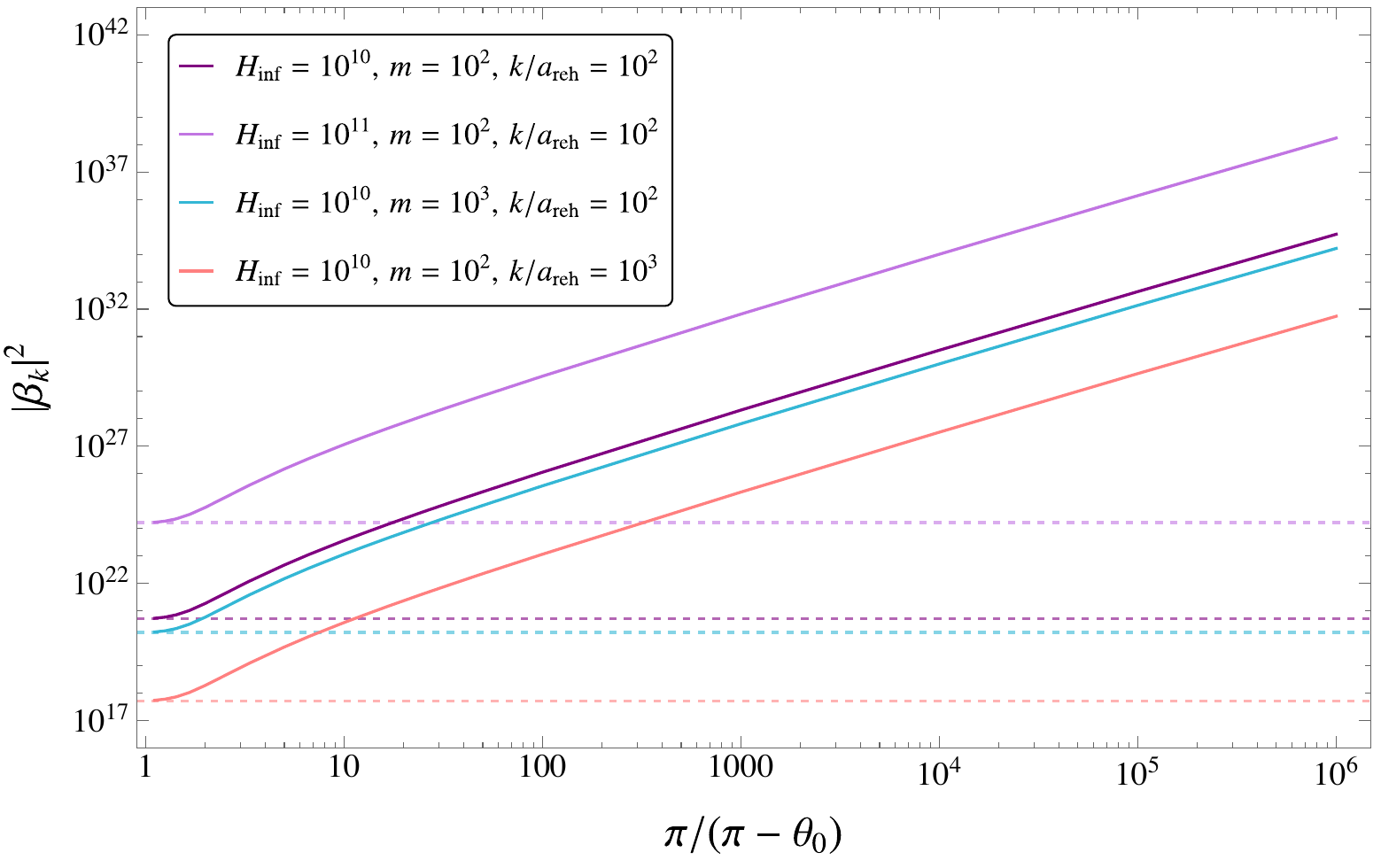}
    \caption{Asymptotic value of the Bogoliubov coefficient~$|\beta_k|^2$ as a function of the axion misalignment angle~$\theta_0$. Colours denote different choices of the inflationary Hubble scale, axion mass, and the wave number upon reheating, as listed in the legend in units of GeV. Dashed horizontal lines denote the offset at $\theta_0 \to 0$, as given in (\ref{betaexpminimum}).}
    \label{betaparameters}
\end{figure}

The expression (\ref{betaexpminimum}) applies for initial field values close to the minimum, such that the potential is well approximated by a quadratic.
To understand the dependence on the parameters for field values where
anharmonic effects are relevant, we have to rely on numerical simulations. The results are shown in Figure \ref{betaparameters}, where the asymptotic value of $\abs{\beta_k}^2$ as a function of the initial misalignment angle is plotted. The different lines correspond to different choices for the Hubble rate, the axion mass, or the wave number; the horizontal dashed lines are the analytical value computed using (\ref{betaexpminimum}).
What we find is that $|\beta_k|^2$ always depends on the same combination of parameters as given in (\ref{betaexpminimum}), which sets the overall normalization for $|\beta_k|^2$. Since the dependence of $|\beta_k|^2$ on $k$ is the same, we can also see that the energy density remains scale independent when the anharmonic effects are taken into account.
Moreover, we see that the shape of $|\beta_k|^2$ as a function of $\theta_0$ is identical for each choice of the parameters. This indicates that the anharmonic enhancement of $|\beta_k|^2$ is determined only by the initial angle.

The final value for $|\beta_k|^2$ also allows us to compute the number of axions today. From (\ref{meanH}) (or (\ref{eq:rho_2-vev})), the physical number density of axion particles that were quantum mechanically produced in the early universe is 
\begin{equation}
 n = \int \frac{dk}{k}\, n_k\,,
\qquad
 n_k = \frac{k^3 \abs{\beta_k}^2}{2 \pi^2 a^3}\,,
\end{equation}
where $n_k$ denotes the number density of particles with comoving momentum $\sim k$. 
(It should be noted that the total number of axions is dominated by the zero mode axions classically produced from the vacuum misalignment.)
Using the expression~(\ref{betaexpminimum}) for axions produced close to the potential minimum, and rewriting the redshift at instantaneous reheating in terms of the inflation scale as
\begin{equation}
 \frac{a_{\ro{today}}}{a_{\ro{reh}}}
\sim 10^{28} 
\left( \frac{H_{\ro{inf}}}{10^{12}\, \ro{GeV}} \right)^{1/2}\,,
\end{equation}
the number density today is obtained as
\begin{equation}
 n_{k \, \ro{today}} \sim 100\, \ro{cm}^{-3}
\left( \frac{m}{1\, \ro{eV}} \right)^{-1/2}
\left( \frac{H_{\ro{inf}}}{10^{12}\, \ro{GeV}} \right)^{2}\,.
\end{equation}

\subsubsection{Alternative Calculation for $|\beta_k|^2$}
\label{betaEnDens}

In order to understand the physics underlying the expression (\ref{betaexpminimum}), let us present an alternative derivation of the asymptotic value of $\abs{\beta_k}^2$ using the energy density of the axion field:
\begin{equation}
 \rho = \frac{(\phi')^2+ \partial_i \phi  \partial_i \phi }{2 a^2}
+ V(\phi)\,.
\end{equation}
Expanding this in terms of the field fluctuation~$\delta \phi$, then the expectation value of the density operator in the initial vacuum defined above (\ref{meanH}) is written as
\begin{equation}
 \langle \rho \rangle
=  \bar{\rho} + 
\langle \rho_2 \rangle
+ \sum_{n = 3}^{\infty}
\frac{d^n V(\bar{\phi})}{d \bar{\phi}^n} \frac{\langle \delta \phi^n \rangle}{n!}\,.
\label{eq:rho_vev}
\end{equation}
Here $\bar{\rho}$ is the energy density of the background field as given in (\ref{rhoback}), and we considered terms linear in $\delta \phi$ to have vanishing expectation values. The quadratic-order density operator~$\rho_2$ is given by
\begin{equation}
 \rho_2 = \frac{1}{2 a^2}
\left\{
(\delta \phi')^2 
+ (\partial_i \delta \phi ) (\partial_i \delta \phi )
+ a^2 m_{\ro{eff}}^2 \delta \phi^2
\right\}\,,
\end{equation}
whose vacuum expectation value is written in terms of the Bogoliubov coefficients as
\begin{multline}
 \langle \rho_2 \rangle
= \frac{1}{2 a^4} \int \frac{d^3 k}{(2 \pi)^3}
\biggl[
\abs{\omega_k}
\left( \abs{\beta_k}^2 + \frac{1}{2} - \frac{\alpha_k \beta_k + \alpha_k^* \beta_k^*}{2} \right)
\\
+ 
\frac{k^2 + a^2 (m_{\ro{eff}}^2 + H^2)}{\abs{\omega_k}}
\left( \abs{\beta_k}^2 + \frac{1}{2} + \frac{\alpha_k \beta_k + \alpha_k^* \beta_k^*}{2} \right)
+ i a H (\alpha_k \beta_k - \alpha_k^* \beta_k^*)
\biggr]\,.
\label{eq:5.5}
\end{multline}
After the onset of the axion oscillations the effective mass asymptotes to the mass around the minimum, $m_{\ro{eff}} \simeq m$ ($ \gg H, k/a$), 
and the frequency to $\abs{\omega_k} \simeq a m $.
In this epoch (\ref{eq:5.5}) takes the form of
\begin{equation}
 \langle \rho_2 \rangle \simeq
\frac{m}{a^3} \int \frac{d^3k}{(2 \pi)^3} 
\left( \abs{\beta_k}^2 + \frac{1}{2} \right)\,.
\label{eq:rho_2-vev}
\end{equation}
This is equivalent to the asymptotic value of the Hamiltonian density~$\langle \mathcal{H} \rangle / a^4 V$, cf.~(\ref{meanH}).

We can also compute the axion density by treating it as a purely classical quantity. For this, notice that the axion field during inflation follows a slow-roll attractor where the field velocity is set by the field value as $\phi' \propto - d V / d \phi$. Hence the asymptotic energy density can be considered to be uniquely determined by the axion field value at some initial time~$\tau_i$ during inflation, i.e., $\rho  = \rho (\phi_i)$. Then the spatial inhomogeneity of the asymptotic density can be understood to originate from that of the initial field, $\phi_i (\bd{x}) = \bar{\phi}_i + \delta \phi_i (\bd{x})$, where $\bar{\phi}_i$ is the average field value over the entire universe at $\tau_i$, and $\delta \phi_i $ is the classical fluctuation around the average. (We are being somewhat sloppy and using the same notation as we used in (\ref{eq:split}) for splitting between the classical background and quantum fluctuations, but we stress that here everything is classical.) Expanding the asymptotic density in terms of $\delta \phi_i (\bd{x})$, and also taking a spatial average yields
\begin{equation}
 \langle \rho (\phi_i ) \rangle_{\ro{avg}} = 
 \rho (\bar{\phi}_i)
+ \frac{\partial^2 \rho(\bar{\phi}_i)}{\partial \bar{\phi}_i^2} \frac{\langle \delta \phi_i^2 \rangle_{\ro{avg}}}{2}
+ \sum_{n = 3}^{\infty}
\frac{\partial^n \rho(\bar{\phi}_i)}{\partial \bar{\phi}_i^n} 
\frac{ \langle \delta \phi_i^n \rangle_{\ro{avg}}}{n!}\,.
\label{eq:rho_avg}
\end{equation}
By identifying the spatially averaged quantities with the vacuum expectation values of the corresponding operators, we can equate (\ref{eq:rho_avg}) with (\ref{eq:rho_vev}). Moreover, the variance $\langle \delta \phi^2 \rangle_{\ro{avg}}$ can be evaluated as the vacuum expectation value of the quadratic operator, as shown in (\ref{PSdeltaphi}).
Hence the second term of (\ref{eq:rho_avg}) is rewritten as
\begin{equation}
 \frac{d^2 \rho(\bar{\phi}_i)}{d \bar{\phi}_i^2} \frac{\langle \delta \phi_i^2 \rangle_{\ro{avg}}}{2}
\simeq
\frac{1}{2}
 \frac{d^2 \rho(\bar{\phi}_i)}{d \bar{\phi}_i^2} 
\int_{k \, \leq \, a_i H_{\ro{inf}} } \frac{d^3 k}{(2 \pi)^3} \frac{H_{\ro{inf}}^2}{2 k^3}\,.
\label{eq:rho-2-der}
\end{equation}
Here we ignored modes that are sub-horizon at the time~$\tau_i$, since the contributions from such modes should be renormalized and absorbed into the cosmological constant. Additionally, for the mode function during inflation, we substituted its value in the super-horizon limit for an effectively massless field, 
$\abs{u_k}^2 / a^2 \simeq H_{\ro{inf}}^2 / 2 k^3$ (cf. $ k \ll a H_{\ro{inf}}$ limit of (\ref{eq:uk_inf}) with $\nu = 3/2$).

Upon comparing (\ref{eq:rho_vev}) and (\ref{eq:rho_avg}), let us identify $\bar{\rho}$ with $\rho (\bar{\phi}_i)$, and further ignore the higher-order correlators. Then we can equate the second terms of each expression, namely, (\ref{eq:rho_2-vev}) and (\ref{eq:rho-2-der}). Let us further suppose that this equality holds mode by mode, and also drop the $1/2$ in (\ref{eq:rho_2-vev}).
Then we find a relation between the asymptotic value of the Bogoliubov coefficient, and the second-order derivative of the background energy density in terms of the initial field value, 
\begin{equation}
 \abs{\beta_k}^2 \sim
\frac{a^3 H_{\ro{inf}}^2}{4 m k^3} 
 \frac{\partial^2 \rho(\bar{\phi}_i)}{\partial \bar{\phi}_i^2}\, .
\label{eq:star}
\end{equation}
Since the energy density after the axion started oscillating redshifts as $\rho \propto a^{-3}$, the right-hand side is actually time-independent.

If the initial field value is close to the potential minimum, i.e. 
$\abs{\bar{\phi}_i} \ll f$, then the axion potential is approximately quadratic and the field begins a harmonic oscillation when $H \sim m$. Hence the asymptotic axion density can approximately be written as
\begin{equation}
 \rho (\bar{\phi}_i) \sim \frac{1}{2} m^2 \bar{\phi}_i^2 
\left( \frac{a_{\ro{osc}}}{a} \right)^3
\sim
\frac{1}{2}  m^{1/2} H_{\ro{inf}}^{3/2} \, \bar{\phi}_i^2 
\left(\frac{a_{\ro{reh}}}{a}\right)^3\,,
\end{equation}
where upon moving to the far right-hand side we used
$a_{\ro{osc}} \sim a_{\rm{reh}} (H_{\ro{inf}} / m)^{1/2} $.
Plugging this into (\ref{eq:star}), 
we can estimate the Bogoliubov coefficient around the minimum as
\begin{equation}
 \abs{\beta_k}^2 \sim
\frac{1}{4}
\left( \frac{a_{\ro{reh}}}{k} \right)^3
\frac{ H_{\ro{inf}}^{7/2}}{m^{1/2}}\,.
\end{equation}
This matches with the exact result~(\ref{betaexpminimum}) up to a numerical factor of order unity.\footnote{The numerical factor of (\ref{betaexpminimum}) can also be obtained from (\ref{eq:star}) by using the exact result for the asymptotic axion density~\cite{Kobayashi:2017jcf} 
(see their Eq.~(A3), and combine with $H = H_{\ro{inf}} (a_{\ro{reh}}/ a)^2$): 
\begin{equation}
 \rho =   \frac{[\Gamma(\frac{1}{4})]^2}{4 \pi }
    m^{\frac{1}{2}}H_{\ro{inf}}^{\frac{3}{2}} \, \bar{\phi}_i^2 
\left(\frac{a_{\ro{reh}}}{a}\right)^3.
\end{equation}
}

We have numerically computed both sides of (\ref{eq:star}); these are displayed as functions of the initial angle\footnote{The background field $\bar{\phi}_i$ in (\ref{eq:star}) denotes the value after the wave mode of interest has left the horizon (cf. discussions below (\ref{eq:rho-2-der})), while in the plots we use $\theta_0 = \bar{\phi}_0 / f$ which is the value when the mode is still deep inside the horizon. However we do not distinguish between $\bar{\phi}_i$ and $\bar{\phi}_0$, since the axion background is effectively frozen during inflation for our choice of parameters.} 
in Figure \ref{betacondiniall}, where the purple line shows the left-hand side, while the pink line shows the right-hand side.
At $\theta_0 \lesssim 1$, where the potential can be approximated by a quadratic, the two results are in good agreement.
In particular, both asymptote to the value of (\ref{betaexpminimum}) as $\theta_0 \to 0$. 
Close to the hilltop, on the other hand, $\abs{\beta_k}^2$ is consistently larger than the right-hand side of (\ref{eq:star}) by an order of magnitude.
We have also checked that the $\theta_0$-dependences of either sides of (\ref{eq:star}) are unaltered by changing the values of $m$, $H_{\rm{inf}}$, and $k$; hence the behaviours discussed above are also parameter-independent.

We should remark that our calculation incorporates axion self-interactions through the interaction between the homogeneous background field (zero mode) and the fluctuation mode, however we have ignored mode-mode interactions by expanding the action only up to quadratic order in the fluctuation. The relation~(\ref{eq:star}) has also been derived by neglecting higher-order correlators. One would naively expect the mode-mode interactions to be suppressed by a hierarchical parameter choice such that $m \ll H_{\ro{inf}} \ll f$. However the uniform discrepancy between both sides of~(\ref{eq:star}) at $\theta_0 \gtrsim 1$ may be implying the importance of the interactions between the fluctuation modes. We leave a detailed investigation of this for the future.

\subsection{Squeezing Parameters}
\label{squeezeevolution}
\begin{figure}[t]
    \centering
    \begin{subfigure}[t]{0.48\textwidth}
        \includegraphics[width=\textwidth]{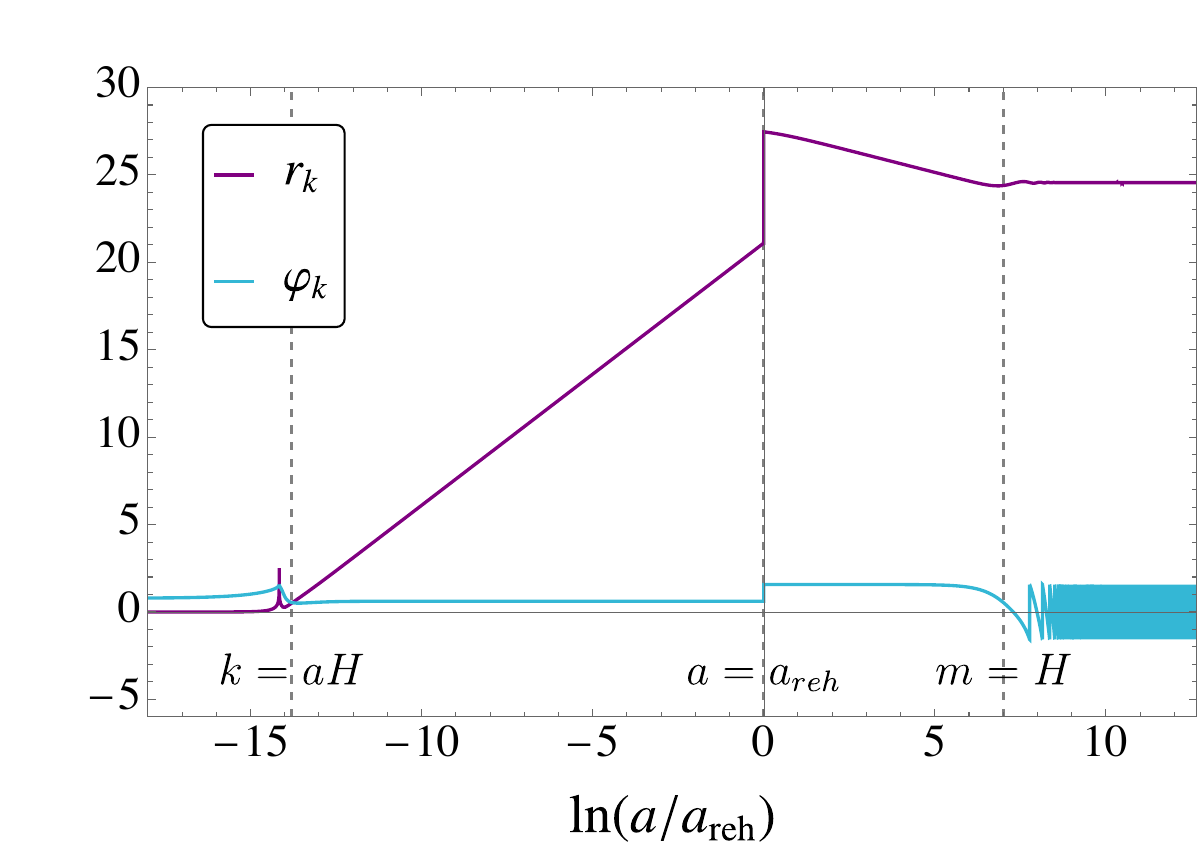}
        \caption{$\theta_0=0.1\,\pi$}
        \label{fig:sq1}
    \end{subfigure}
    \quad
    \begin{subfigure}[t]{0.48\textwidth}
        \includegraphics[width=\textwidth]{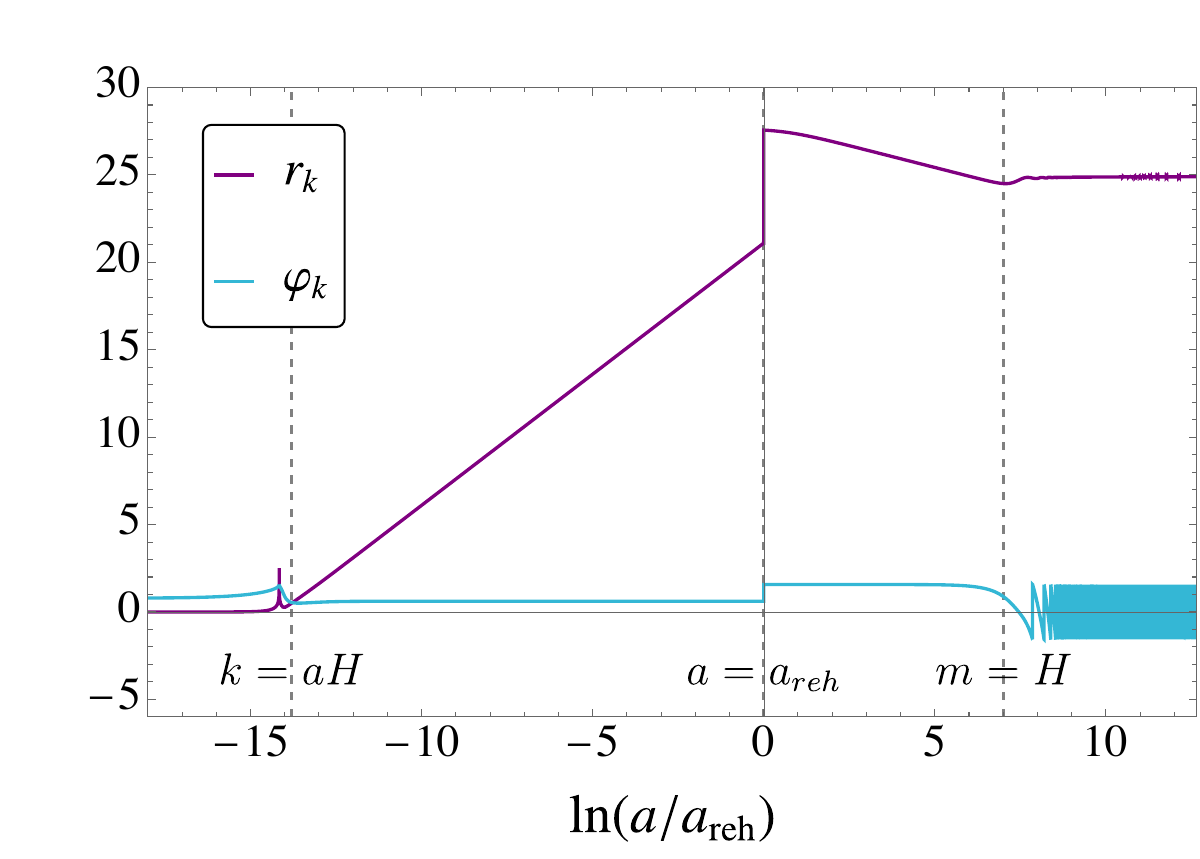}
        \caption{$\theta_0=0.4\,\pi$}
        \label{fig:sq2}
    \end{subfigure}
    
    \begin{subfigure}[t]{0.48\textwidth}
        \includegraphics[width=\textwidth]{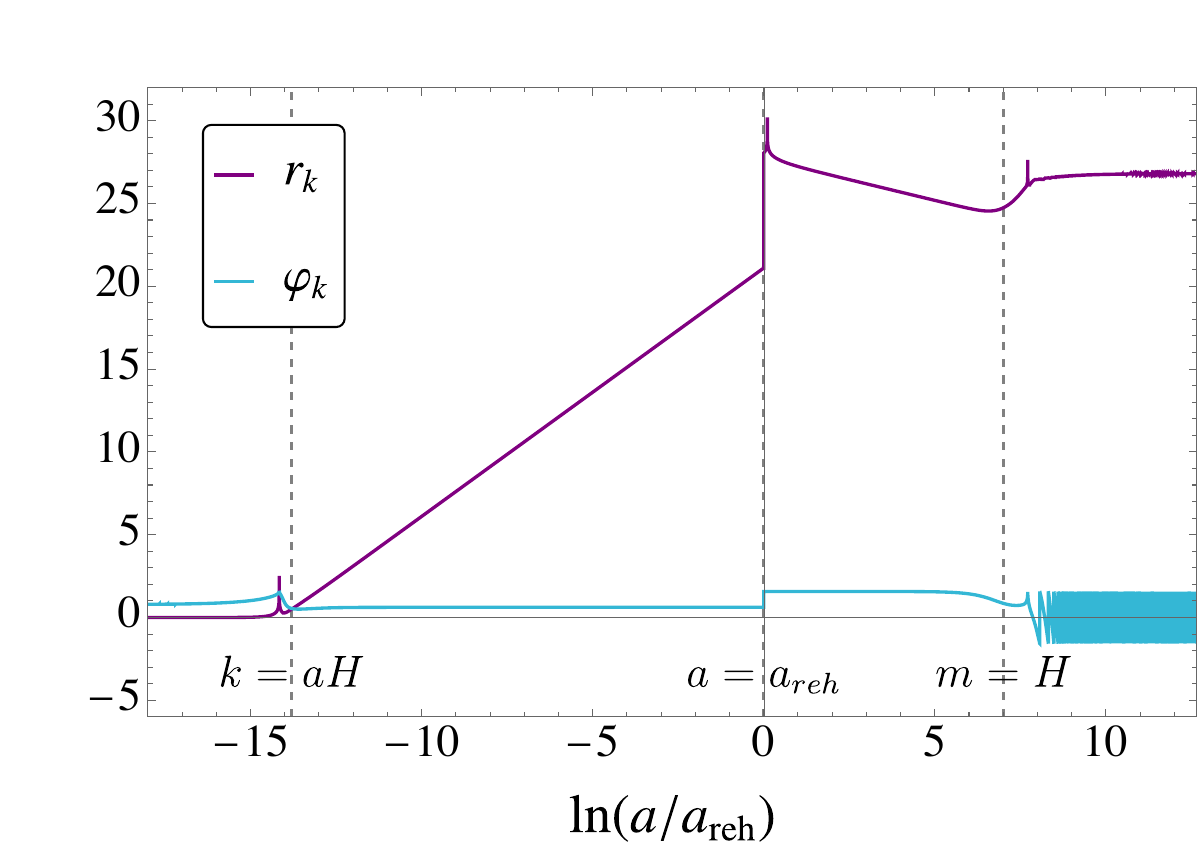}
        \caption{$\theta_0=0.8\,\pi$}
        \label{fig:sq3}
    \end{subfigure}
    \quad
    \begin{subfigure}[t]{0.48\textwidth}
        \includegraphics[width=\textwidth]{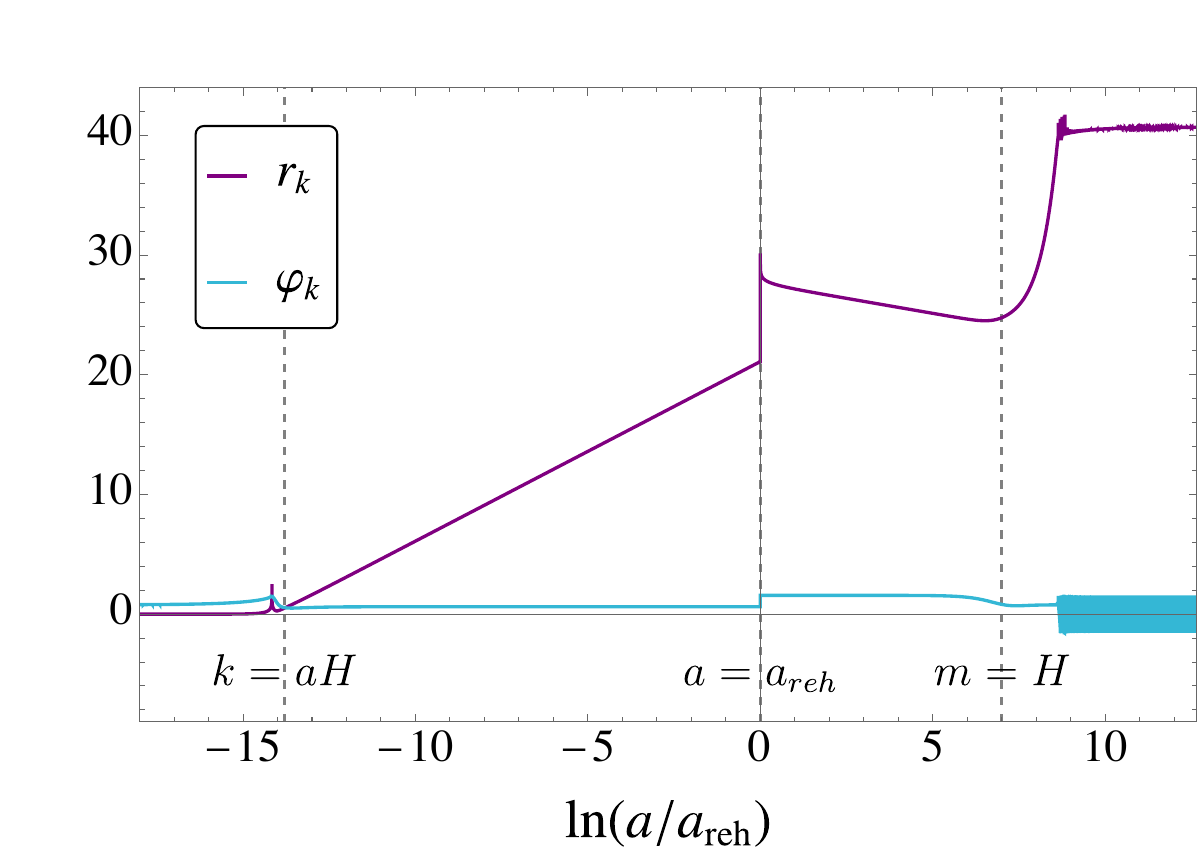}
        \caption{$\theta_0=0.999999\,\pi$}
        \label{fig:sq4}
    \end{subfigure}
    \caption{Time evolution of the squeezing parameters $r_k$ and $\varphi_k$ for different values of the initial misalignment angle. The parameters chosen for the evaluation are $H_\ro{inf}=10^8\, \text{GeV}$, $m=10^2\,\text{GeV}$, and $k/a_\ro{reh}=10^2\,\text{GeV}$. The dashed vertical lines set the horizon exit ($k=aH$), reheating ($a=a_\ro{reh}$) and the onset of the oscillations of the background field ($m=H$).}
    \label{squeezingplots}
\end{figure}
We can now proceed to the analysis of the squeezing parameters. Through relations (\ref{sqparf}), one gets the evolution in time of the squeezing parameters. The result is shown in Figure~\ref{squeezingplots} for $r_k$ and $\varphi_k$, which are the two parameters linked with physical quantities, as explained in Section \ref{squeezingparameters} and in Appendix \ref{sqparphysmean}. Their evolution is reported for four different values of the initial misalignment angle.

The $r_k$ parameter is directly linked with the modulus square of the beta coefficient; we can therefore see an enhancement in its value towards the hilltop. $r_k$ gives a measure of the stretching of the ellipse describing the phase space distribution of two conjugate variables, as explained in Appendix~\ref{sqparphysmean}. If $r_k$ increases, as observed taking into account the anharmonic effects, then the ellipse gets stretched along one of its axis, and the state described by the two conjugate variables becomes more and more squeezed. We therefore conclude that anharmonic effects give extra squeezing at the onset of the oscillations.

The angle $\varphi_k$ is instead linked with the rotation of the ellipse in phase space. 
$\varphi_k$ is constant until the mode exits the horizon, then shifts to another constant value after the horizon exit.
When the background field starts oscillating, and the adiabaticity condition starts to hold, the phase starts to rotate. This behaviour is visible in Figure \ref{squeezingplots}. The sharp lines are present since the phase is restricted to the range
$-\pi \leq \varphi_k \leq \pi$. In phase space, the ellipse described by this phase is constantly rotating by this time dependent angle.

\section{Conclusions}
\label{conclusions}

We have computed the time evolution of the Bogoliubov coefficients and the squeezing parameters for an axion-like field during inflation and 
radiation domination, keeping into account self-interactions. In particular we considered the non-trivial evolution of the background field.
We have shown that the net number of particles created due to the expansion of the Universe, given by $|\beta_k|^2$, increases as the initial background field approaches the hilltop of the cosine potential. The $r_k$ parameter determining the amount of squeezing increases accordingly. Hence we conclude that the self-interaction due to the cosine potential, and specifically the interaction between the perturbation and the non-trivial background field dynamics, gives rise to extra post-inflation squeezing.

An analytical expression for $|\beta_k|^2$ in terms of the relevant parameters of the system has also been found in the quadratic potential limit in (\ref{betaexpminimum}). A full derivation has been carried out in terms of the analytical mode functions in Appendix~\ref{quadraticmodes}. The expression (\ref{betaexpminimum}) has also been confirmed via numerical computations to set the overall normalization of $|\beta_k|^2$. The behaviour of the final value of $|\beta_k|^2$ as a function of the initial displacement angle is not affected by the physical parameters $H_\ro{inf}$, $m$ and $k$, provided the mode considered re-enters the horizon after the background has started oscillating.
Moreover the anharmonic enhancement of the squeezing is solely determined by the initial misalignment angle.

We have also derived a relation between the asymptotic values of the Bogoliubov coefficients, and the field derivatives of the background energy density, as shown in (\ref{eq:star}). This relation exactly holds for small misalignment angles where the axion potential is well approximated by a quadratic, however we also found that it breaks down towards the hilltop. Here we note that in this paper we evaluated the effect of axion self-interactions through the interaction between the homogeneous background field and the fluctuation, while we have ignored interactions among the fluctuation modes. The breakdown of (\ref{eq:star}) towards the hilltop may be indicating that mode-mode couplings can further enhance axion squeezing. We leave an investigation of this effect for future work.

A connection with observables has also been pointed out. The net number of particles created by the expansion of the Universe is given by $|\beta_k|^2$, which is directly linked also to the parameter $r_k$ and consequently to the amount of squeezing of the perturbations. The more particles are created, the more squeezed the state is. The other physically relevant squeezing parameter, i.e. the phase $\varphi_k$, is related with the power spectrum of the field fluctuations, as shown in (\ref{powersq}).

One interesting question left for future work is how we could observationally discern among the quantum or classical origin of the axion perturbations. The main problem in addressing this question is the fact that the two-point correlation functions of a squeezed state can be perfectly mimicked by a stochastic classical distribution. Despite this, a squeezed state is an intrinsically quantum state and in particular it is an entangled state.
Hence one possible test of the quantum nature of the perturbations is a Bell type experiment on the Cosmic Microwave Background (CMB). In the case of the curvature perturbations this kind of experiment is however inconclusive, see \cite{Martin:2017zxs} and references therein. Even if considering Leggett-Garg inequalities \cite{Emary:2013wfl} instead of Bell inequalities, the curvature perturbations could be useless, since the conjugate momentum of the curvature perturbation $\zeta$, i.e. its time derivative, is vanishing on super-horizon scales. Working with axion fields may avoid this latter problem. We have indeed shown that the phase $\varphi_k$ oscillates in the late time limit, making the ellipse in phase space describing the system to rotate. This suggests that the time derivative of the perturbation is non-negligible, and hence axions may provide an ideal target for a cosmological Leggett-Garg experiment.

Another way to probe the nature of the primordial fluctuations is to consider interacting theories and compute higher order correlators. One attempt in this regard have been performed in \cite{Green:2020whw} (see also in a different context, e.g., \cite{Martin:2018lin, DaddiHammou:2022itk}). In the recent work by Green and Porto \cite{Green:2020whw} it has been pointed out how the quantum origin of the perturbations could be singled out by looking at the poles of the bispectrum, therefore a careful analysis of the axion bispectrum might be enlightening. In order to compute the bispectrum of an axion field, the effect of the non-trivial background computed in this work will have to be taken into account.

Although we have focused on axion fields throughout this paper, the study of the squeezing of the perturbations can also be important for addressing the evolution of gauge fields that could have been excited in the early Universe. Recently a paper has been published on this topic \cite{Tripathy:2023aha} and further analysis along this direction is an interesting expansion of our work.

\section*{Acknowledgements}
We thank Angelo Ricciardone and Marco Peloso for useful discussions.
T.K. acknowledges support from the INFN program on Theoretical Astroparticle Physics (TAsP), and JSPS KAKENHI (Grant No. JP22K03595). N.B., V.D., S.M. and M.V. are supported by the INFN INDARK grant. N.B. and S.M. thank partial financial support for this work by the MUR Departments of Excellence grant 2023-2027 “Quantum Frontiers”. MV is also supported by the ``Italian Research Center on High-Performance Computing, Big Data and Quantum Computing'' funded by MUR Missione 4 Componente 2 Investimento 1.4 - Next Generation EU (NGEU).

\appendix

\section{Physical Interpretation of the Squeezing Parameters}
\label{sqparphysmean}

The squeezing formalism is an alternative way to describe the process of particle creation. Its advantage is to give a clear phase space representation of the system's evolution. Introductions to the squeezing formalism can be found in \cite{Hu:1993gm,Calzetta:2008iqa} (see also \cite{Martin:2007bw, Martin:2015qta, Martin:2021znx, Martin:2022kph}). The evolution in time of the ladder operators can be given in terms of an evolution operator~$U$ as
\begin{equation}
\label{aevolv}
    a_{ \boldsymbol{k}}(\tau) = U(\tau)\, a^{0}_{\boldsymbol{k}}\, U^\dagger(\tau)\,, 
\qquad
    a^{\dagger}_{ \boldsymbol{k}}(\tau) = U(\tau)\, a^{0 \dagger}_{\boldsymbol{k}}\, U^\dagger(\tau)\,.
\end{equation}
The form of this operator has been deduced by Parker in \cite{Parker:1969au}. Using the language introduced by \cite{Caves:1985zz} in the context of quantum optics, it can be rewritten as $U=RS$, where
\begin{equation}
\label{Rrotation}
    R(\vartheta_{k}) = \exp\left[ i \int d^3k\,  \vartheta_k a_{\bd{k}}^{0 \dagger} a_{\bd{k}}^0 \right]
\end{equation}
is the rotation operator, and
\begin{equation}
\label{Ssqueezing}
    S(r_k,\varphi_k) = \exp\left[ \frac{1}{2} \int d^3k \, r_k \left( e^{-2 i \varphi_k} a_{\bd{k}}^0 a_{-\bd{k}}^0 - e^{2 i \varphi_k} a_{\bd{k}}^{0 \dagger} a_{-\bd{k}}^{0 \dagger} \right)\right]
\end{equation}
is the two-mode squeeze operator. Here $r_k$, $\varphi_k$ and $\vartheta_k$ are the squeezing parameters introduced in~(\ref{sqpar}).
Their time dependence in (\ref{Rrotation}) and (\ref{Ssqueezing}) have been made implicit to avoid clutter.
The time evolution (\ref{aevolv}) thus corresponds to the application of a rotation with the parameter $\vartheta_k$, followed by a squeezing with the parameters $r_k$ and $\varphi_k$. The action of these two operators can be computed using the expansion $e^{\hat A}\,\hat B\, e^{-\hat A} = \hat B + [\hat A, \hat B] + 1/2\, [\hat A, [\hat A, \hat B]] + \dots$ as
\begin{equation}
\begin{split}
 \label{aevolv1}
   & U a_{\boldsymbol{k}}^0\, U^\dagger = e^{-i\,\vartheta_k}\, \cosh r_k \, a_{\boldsymbol{k}}^0 + e^{i\,\left(\vartheta_k + 2\,\varphi_k\right)}\, \sinh r_k \, a_{-\boldsymbol{k}}^{0\,\dagger}\,,
\\
& U a_{\boldsymbol{k}}^{0 \, \dagger}\, U^\dagger = e^{i\,\vartheta_k}\, \cosh r_k \, a_{\boldsymbol{k}}^{0 \, \dagger} + e^{-i\,\left(\vartheta_k + 2\,\varphi_k\right)}\, \sinh r_k \, a_{-\boldsymbol{k}}^{0}\,.
\end{split}
\end{equation}
Rewriting the squeezing parameters in terms of the Bogoliubov coefficients using (\ref{sqparf}), we see that (\ref{aevolv1}) matches with (\ref{bogo}). 
These expressions also show that the modes $+\boldsymbol{k}$ and $-\boldsymbol{k}$ are mixed, hence a two-mode squeeze operator is needed to describe particle pairs in states $\pm \boldsymbol{k}$.

The rotation operator shifts the phase of the ladder operators as
$R a_{\boldsymbol{k}}^0\, R^\dagger = e^{-i \vartheta_{k}} a_{\boldsymbol{k}}^0$, 
$R a_{\boldsymbol{k}}^{0\, \dagger}\, R^\dagger = e^{i \vartheta_{k}} a_{\boldsymbol{k}}^{0\, \dagger}$.
The vacuum state is invariant under this rotation \cite{Calzetta:2008iqa, Martin:2007bw}, and hence the parameter $\vartheta_k$ is unphysical.
Physical observables thus depend only on the parameters $r_k$ and $\varphi_k$. Using (\ref{sqparf}), this in turn implies that the relevant combinations of the Bogoliubov coefficients are $|\beta_k|^2$ and $\arg \left( \alpha_k \beta_k \right)$. In particular, the squeezing parameter $r_k$ is related to the number of particles created. 
The expectation value of the number operator in the vacuum annihilated by $a_{\boldsymbol{k}}^0$ follows from (\ref{aevolv1}) as,
\begin{equation}
\label{betasqparr}
   \frac{(2\pi)^3}{V}
 \langle  a^\dagger_{\boldsymbol{k}} a_{\boldsymbol{k}} \rangle = \sinh^{2} r_k = |\beta_k|^2,
\end{equation}
as we have also derived in (\ref{eq:6.18}), and $V$ is the spatial volume. 
Thus cosmological particle creation amounts to squeezing the vacuum, as observed by Grishchuk and Sidorov \cite{Grishchuk:1990bj}.
\begin{figure}[t]
    \centering
    \includegraphics[scale=.7]{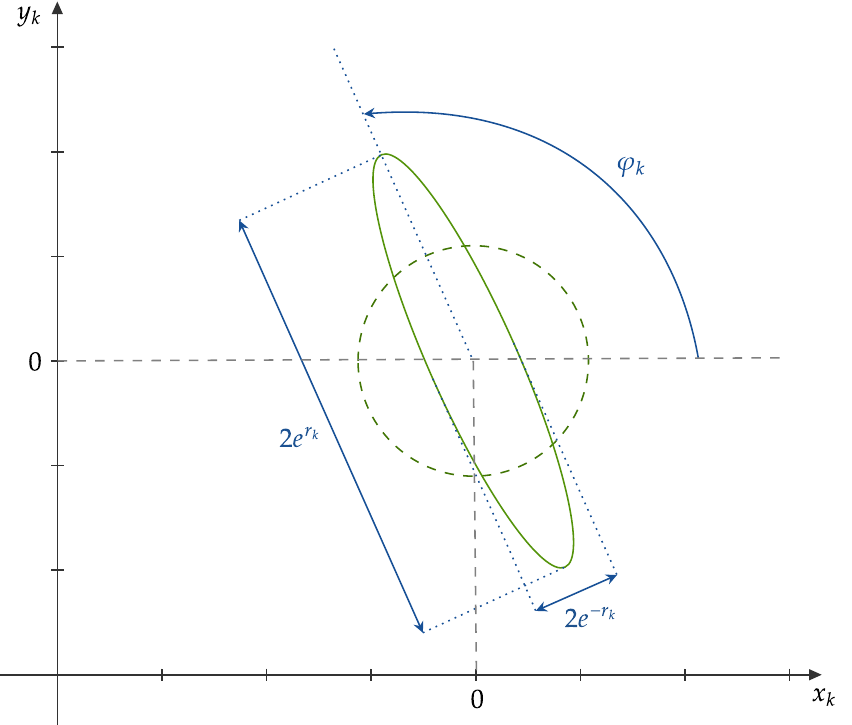}
    \caption{Phase space distribution of a conjugate pair $(x_k,y_k)$. The green circle corresponds to a vacuum state, while the green ellipse is a squeezed state described by the parameter $r_k$ and rotation angle $\varphi_k$. This picture has been adapted from \cite{phdthesis}.}
    \label{fig:sqphasespace}
\end{figure}

In order to understand better the physical meaning of the squeezing parameters we need to move to the phase space representation for a quadrature pair.
A squeezed state is described by a quadrature pair for which fluctuations in one variable are reduced below the symmetric quantum limit, while fluctuations in the other canonically conjugate quadrature component are enhanced so that the Heisenberg uncertainty principle is not violated.
To see this explicitly, we introduce a set of operators:
\begin{equation}
x_{\boldsymbol{k}} = \frac{1}{\sqrt{2}} \left( a_{\boldsymbol{k}} + a^\dagger_{-\boldsymbol{k}} \right),
\quad
y_{\boldsymbol{k}} = - \frac{i}{\sqrt{2}} \left( a_{\boldsymbol{k}} - a^\dagger_{-\boldsymbol{k}}  \right),
\end{equation}
which are the same as $\chi_{\bd{k}}$ and $p_{\bd{k}}$ in the main text, up to the normalization factor~$\sqrt{\abs{\omega_k}}$.
These satisfy the commutation relation,
\begin{equation}
 \left[ x_{\boldsymbol{k}}, y^{\dagger}_{\boldsymbol{k}^\prime} \right] = i \delta^{(3)}(\boldsymbol{k}-\boldsymbol{k}^\prime).
\label{eq:xy-com}
\end{equation}
We further introduce a linear combination of the operators by rotating 
in the $x_{\bd{k}}$ - $y_{\bd{k}}$ plane by an angle~$\psi$~($\in \mathbb{R}$),
\begin{equation}
 \xi_{\bd{k}} (\psi) = \cos \psi \, x_{\bd{k}} + \sin \psi \, y_{\bd{k}}.
\end{equation}
The vacuum expectation value of a quadratic combination of $\xi_{\bd{k}}$ can be written as,
\begin{equation}
\left|
 \frac{\langle \xi_{\bd{k}} (\psi) \xi^{\dagger}_{\bd{k}} (\psi) \rangle}{ \frac{1}{2} [x_{\bd{k}}, \, y^{\dagger}_{\bd{k}} ] }
\right|^{1/2} 
= \left\{
\cosh (2 r_k) + \cos (2 \psi - 2 \varphi_k) \sinh (2 r_k)
\right\}^{1/2}.
\label{eq:A.9}
\end{equation}
The right-hand side takes the values 
$e^{r_k} $ for $\psi =  \varphi_k$ and
$e^{-r_k} $ for $\psi =  \varphi_k + \frac{\pi }{2}$, 
moreover for $r_k = 0$ it becomes unity for all~$\psi$.
The quantity of~(\ref{eq:A.9}) denotes the standard deviation of $x_{\bd{k}}$ and $y_{\bd{k}}$ along the $\psi$-direction, in units of half the quantum uncertainty~(\ref{eq:xy-com}).
It also implies that the probability distribution in the $x_{\bd{k}}$ - $y_{\bd{k}}$ plane is an ellipse with 
semi-major axis~$e^{r_k}$ and semi-minor axis~$e^{-r_k}$ (normalized by the quantum uncertainty), with the major axis rotated by an angle~$\varphi_k$, as illustrated in Figure~\ref{fig:sqphasespace}.
This can be understood by noting that when projecting this ellipse onto the $\psi$-axis, then its half length is given by the
right-hand side of~(\ref{eq:A.9}).\footnote{The ellipse in an $x$ - $y$ plane is described by 
\begin{equation}
 e^{-2 r_k} (\cos \varphi_k \, x + \sin \varphi_k \, y )^2
+  e^{2 r_k} ( -\sin \varphi_k \, x + \cos \varphi_k \,  y )^2
 = 1.
\label{eq:A10}
\end{equation}
To obtain the length of this ellipse projected along the $\psi$-axis, consider a line running through the origin with a polar angle~$\psi$, and another line that is orthogonal to the first line and passes through a point $(\rho \cos \psi, \rho \sin \psi)$ with $\rho > 0$, i.e., 
\begin{equation}
 y  = -\frac{x}{\tan \psi} + \frac{\rho }{\sin \psi }.
\label{eq:A11}
\end{equation}
This line is tangent to the ellipse if (\ref{eq:A10}) and (\ref{eq:A11}) admit only one set of solutions for ($x, y$). 
This requirement sets~$\rho$ as the projected half length, and reproduces the right-hand side of~(\ref{eq:A.9}).}
When $r_k = 0$ the ellipse in phase space reduces to a circle; in this case, the state is the vacuum state of the system.
The $\vartheta_k$~angle does not enter the discussion here since we are considering vacuum expectation values.
More details about the phase space representation using the squeezing parameters can be found in~\cite{Albrecht:1992kf,Micheli:2022tld,Martin:2021znx}.

For the axion field fluctuation analyzed in the main text, the parameters $r_k$ and $\varphi_k$ depend on time such that the ellipse in phase space becomes more and more squeezed, and also rotates as the Universe expands. These effects start to apply when the fluctuation's wave mode exits the Hubble horizon during inflation. The squeezing continues until adiabaticitiy is recovered, when the background axion field starts to oscillate. 
In this example, the expanding Universe and the rolling axion background both induce a cosmological squeezing of the axion fluctuations.

\section{Quadratic Potential}
\label{quadratic}

In this appendix we study a quadratic axion potential, and derive analytic solutions for the mode function, as well as the asymptotic values of the Bogoliubov coefficients.
The analytical results presented coincide with the numerical results obtained for the full cosine potential in the limit of a small initial displacement angle, close to the minimum of the potential.

\subsection{Mode Functions}
\label{quadraticmodes}

The equation of motion for the mode function is given by (\ref{modeeq}),
with $m_{\ro{eff}}$ replaced by a constant and positive mass $m$, as here we focus on a quadratic potential.
Let us solve this equation in the dS and RD epochs, assuming the two epochs to be connected by an instantaneous reheating.

\paragraph{dS epoch}
With a constant Hubble rate $H = H_{\ro{inf}}$, the conformal time is related to the scale factor as
\begin{equation}
 \tau = - \frac{1}{a H_{\ro{inf}}}.
\end{equation}
Here we set the time at reheating as $\tau_{\ro{reh}} = - 1/a_{\ro{reh}} H_{\ro{inf}}$.
The equation of motion then reduces to the form,
\begin{equation}
\label{modeeqDS1}
    u_k'' + \left\{
 k^2 + \frac{1}{\tau^2} 
\left( \frac{m^2}{H_{\ro{inf}}^2} - 2\right)
\right\}\,u_k=0\,.
\end{equation}
Supposing $m^2/H_{\ro{inf}}^2 < 9/4$, the solution is given in terms of Hankel functions as,
\begin{equation}
\label{modeDS}
    u_k = (-\tau )^{\frac{1}{2}} \left\{  c_1\, H_{\nu}^{(1)}(-k \tau )+ c_2\, H_{\nu}^{(2)}(-k \tau )\right\} ,
\qquad
 \nu =  \left(\frac{9}{4} - \frac{m^2}{H_{\ro{inf}}^2}\right)^{\frac{1}{2}}.
\end{equation}
The two constants $c_1$ and $c_2$ can be fixed by requiring that the mode function approaches a positive frequency solution in the asymptotic past, and also that it satisfies the normalization condition~(\ref{normalization}).
This yields, up to an unphysical phase, 
\begin{equation}
 u_k = \left(\frac{\pi }{4 a H_{\ro{inf}}}\right)^{\frac{1}{2}}
H_{\nu}^{(1)} \left( \frac{ k }{a H_{\ro{inf}} } \right),
\label{eq:uk_inf}
\end{equation}
where we have rewritten the solution in terms of the scale factor.

\paragraph{RD epoch}
In a radiation-dominated universe where $H \propto a^{-2}$, 
the equation of motion can be rewritten as 
\begin{equation}
 \frac{d^2 u_k}{dz^2} + \left(\frac{z^2}{4} + \frac{\kappa^2}{2 \mu }
\right) u_k = 0,
\end{equation}
where
\begin{equation}
\kappa = \frac{k}{a_{\ro{reh}} H_{\ro{inf}}}\,,
\quad
  \mu = \frac{m}{H_{\ro{inf}}}\,,
\quad
z= (2 \mu )^{\frac{1}{2}} \frac{a}{a_{\ro{reh}}}\,.
\end{equation}
Solutions to this equation are the parabolic cylinder functions,
\begin{equation}
 u_k = c_3 D_{-\lambda } \left( e^{i \frac{\pi}{4}} z \right)
+ c_4 D_{\lambda - 1} \left( e^{i\frac{3\pi}{4}} z \right),
\qquad
 \lambda = \frac{1}{2} + \frac{i \kappa^2}{2\mu }\,.
\label{modeRD}
\end{equation}
The two constants $c_3$ and $c_4$ are fixed by 
imposing continuity of the solutions
(\ref{eq:uk_inf}) and (\ref{modeRD}), as well as their time derivatives, at the time of reheating. 

\subsection{Asymptotic $|\beta_k|^2$}
\label{quadraticexpansion}

In order to evaluate the Bogoliubov coefficient~$|\beta_k|^2$ in the asymptotic future, we introduce linear combinations of parabolic cylinder functions, 
\begin{align}
&\mathcal{F}_+  = 
\frac{e^{-\frac{\pi \kappa^2}{8 \mu} }}{(2 \mu )^{\frac{1}{4}} (a_{\ro{reh}} H_{\ro{inf}})^{\frac{1}{2}} }
D_{-\lambda} 
\left( e^{i \frac{\pi}{4}} z \right)\,,
\\
&\mathcal{F}_-  = 
\frac{e^{\frac{3\pi \kappa^2}{8 \mu} }}{(2 \mu )^{\frac{1}{4}} 
( a_{\ro{reh}} H_{\ro{inf}} )^{\frac{1}{2}} }
\left\{
D_{\lambda - 1} 
\left( e^{i \frac{3\pi}{4}} z \right)
+ \frac{ (2 \pi)^{\frac{1}{2}} e^{-i \frac{3\pi}{4} - \frac{\pi \kappa^2}{4 \mu}} }{\Gamma (- \lambda + 1)}
D_{-\lambda } 
\left( e^{i \frac{\pi}{4}} z \right)
\right\}\,,
\end{align}
and rewrite the mode function in the RD epoch~(\ref{modeRD}) as
\begin{equation}
 u_k  = A_k \, \mathcal{F}_+  + B_k \, \mathcal{F}_- \, .
\label{eq:uk_RD}
\end{equation}
Here $A_k$ and $B_k$ are time-independent coefficients.
The functions $\mathcal{F}_+$ and $\mathcal{F}_-$ are chosen such that in the asymptotic future, they respectively approach the WKB solutions with positive and negative frequencies:
\begin{equation}
 \frac{
e^{\mp i \int^\tau d\tilde{\tau} \, \omega_k (\tilde{\tau})}
}{\sqrt{2 \omega_k (\tau)}} 
\sim 
\frac{e^{\mp i \left( \frac{z^2}{4} + \ro{const.} \right) }}{ (2 \mu )^{\frac{1}{4}} 
(a_{\ro{reh}} H_{\ro{inf}} \, z)^{\frac{1}{2}}}\,.
\label{eq:A.6}
\end{equation}
This can be checked by noting that for an argument~$w$ with a sufficiently large amplitude and $s \neq 0, 1, 2, \cdots$, the parabolic cylinder function is approximated by
\begin{equation}
 D_s (w) \sim e^{-\frac{w^2}{4} } w^s - 
\frac{\sqrt{2 \pi}}{ \Gamma (-s) }e^{i \pi s + \frac{w^2}{4}} w^{-s - 1}\,,
\end{equation}
for $\pi / 4 < \ro{arg} \, w < 5 \pi / 4 $.
For $ \abs{\ro{arg} \,  w} < 3 \pi / 4 $, the approximate expression is given only by the first term in the right hand side.
Hence the amplitudes of the coefficients $A_k$ and $B_k$ in (\ref{eq:uk_RD}) match with those of the Bogoliubov coefficients $\alpha_k$ and $\beta_k$ in the asymptotic future.

By matching the mode functions (\ref{eq:uk_inf}) and (\ref{eq:uk_RD}), as well as their time derivatives~$u_k'$ at $a = a_{\ro{reh}}$, 
the coefficient $B_k$ can be obtained as,
\begin{multline}
 B_k = 
\frac{ \pi^{\frac{1}{2}} e^{i \frac{\pi}{4} -\frac{3 \pi \kappa^2}{8 \mu} } }{2^{\frac{5}{4}} \mu^{\frac{1}{4}} }
\\ \times 
\frac{
\kappa H_{\nu + 1}^{(1)} (\kappa) 
D_{- \lambda} 
- H_{\nu}^{(1)} (\kappa) 
\left\{ 
( \frac{1}{2} + \nu +  i \mu )
D_{- \lambda } 
+ e^{- i \frac{3\pi}{4}} 
\sqrt{2\mu} \, D_{-\lambda + 1} 
\right\}
}{
i D_{-\lambda + 1 } 
D_{\lambda - 1 }  + 
D_{- \lambda } 
\left\{
D_{\lambda } 
+ e^{-i \frac{\pi}{4}} \sqrt{2\mu} D_{\lambda - 1} 
\right\}
}.
\label{eq:fullBk}
\end{multline}
Here the arguments of the parabolic cylinder functions are 
$e^{i \pi/4} \sqrt{2\mu}$ for $D_{-\lambda}$ and $D_{-\lambda + 1}$,
while $e^{i 3\pi/4} \sqrt{2\mu}$
for $D_{\lambda}$ and $D_{\lambda - 1}$.
This expression for $B_k$ can by simplified by first expanding in small~$\kappa$, and then further simplifying the leading term by considering a small~$\mu$.
This gives,\footnote{We use that the Hankel function of the first kind at $w \to 0$ asymptotes to
\begin{equation}
 H_{s}^{(1)} (w) \sim -\frac{i}{\pi }\Gamma (s) 
\left( \frac{w}{2} \right)^{-s}\,,
\end{equation}
for $\ro{Re} (s) > 0$. 
We also use
\begin{equation}
 D_{\frac{1}{2}} (0) = \frac{2^{\frac{1}{4}} \pi^{\frac{1}{2}}}{\Gamma (\frac{1}{4})}\,,
\quad
 D_{-\frac{1}{2}} (0) = \frac{\pi^{\frac{1}{2}}}{ 2^{\frac{1}{4}} \Gamma (\frac{3}{4})}\,.
\end{equation}}
\begin{equation}
 B_k \sim - \frac{i \Gamma (\frac{1}{4})}{2^{\frac{3}{2}} \pi^{\frac{1}{2}} }
\frac{1}{\kappa^{\frac{3}{2}} \mu^{\frac{1}{4}}}\,.
\end{equation}
This gives a good approximation for (\ref{eq:fullBk}) at $\kappa^2 \lesssim \mu \ll 1$; here note that $\kappa^2 \lesssim \mu$ is equivalent to $k \lesssim (a H)_{\ro{osc}}$.
Hence for wave modes that are outside the Hubble horizon when the axion begins to oscillate during the RD epoch, the asymptotic value of the Bogoliubov coefficient is obtained as
\begin{equation}
 \abs{\beta_k}^2 \sim
\frac{[\Gamma(\frac{1}{4})]^2}{8 \pi }
\left( \frac{a_{\ro{reh}}}{k} \right)^3
\frac{ H_{\ro{inf}}^{\frac{7}{2}}}{m^{\frac{1}{2}}}\,.
\end{equation}

\section{More Realistic Models for the Evolution of the Universe}
\label{morerealistic}

In the results derived in the main text, we assumed a pure dS spacetime followed by an RD spacetime, further assuming the reheating phase to be instantaneous. In this Appendix, we check that changing the evolution of the Hubble rate does not affect our main conclusions on the asymptotic value for $|\beta_k|^2$ and the squeezing parameters.

\subsection{Smooth Reheating}
\label{Hsmooth}

\begin{figure}[t]
    \centering
    \includegraphics[scale=.65]{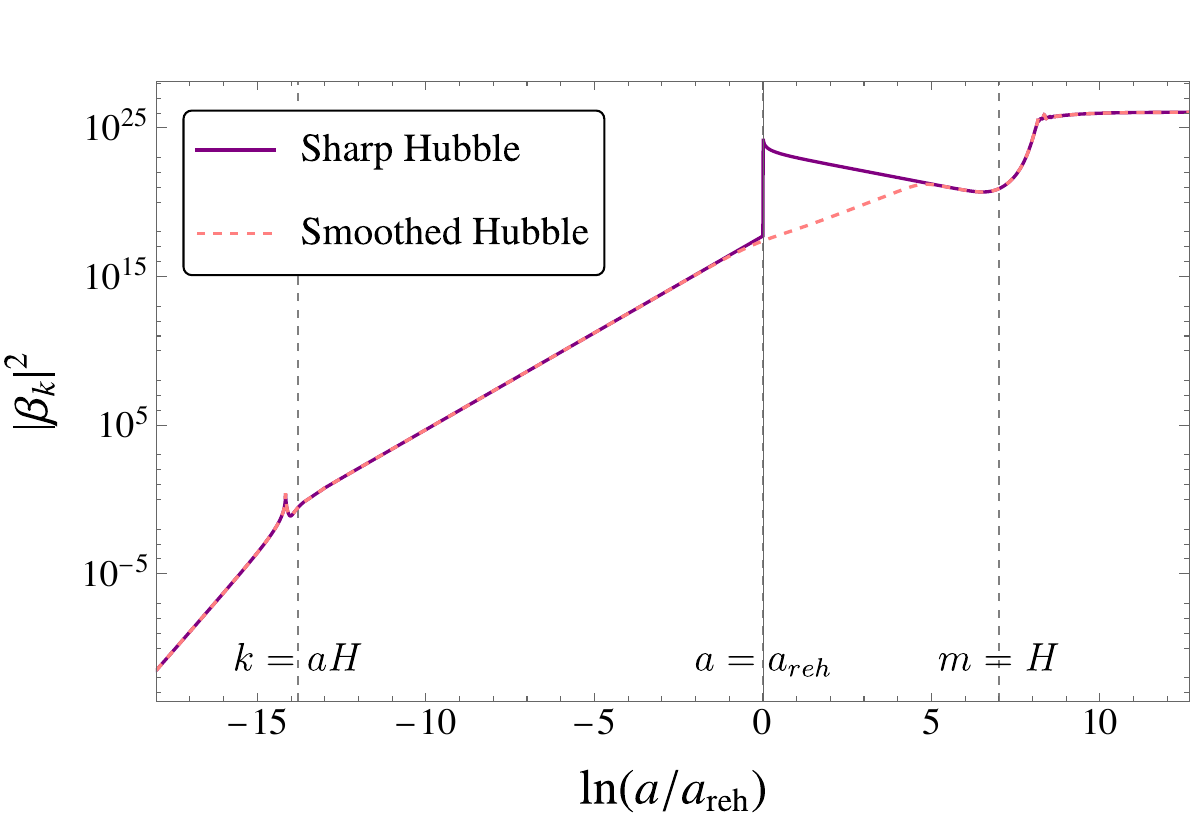}
    \caption{Time evolution of $|\beta_k|^2$ for instantaneous reheating (purple) and smooth reheating (pink dashed), for an initial angle $\theta_0=0.99\,\pi$. The values chosen for the parameters are $H_\ro{inf}=10^8\,\text{GeV}$, $m=10^2\,\text{GeV}$ and $k/a_\ro{reh}=10^2\,\text{GeV}$.}
    \label{betasmooth}
\end{figure}

We first smooth the Hubble rate evolution between the dS and RD epochs to see if our results are affected by a non-instantaneous reheating phase. The expression we use is the following:
\begin{equation}
\label{hubblesmooth}
    H = H_{\ro{inf}}\,\frac{a^{2}_\ro{reh}}{a^{2}+a_\ro{reh}^{2}}
\end{equation}
We numerically solved the axion's equations of motion with the Hubble rate (\ref{hubblesmooth}), and we found that it does not change the evolution of the background field nor the fluctuation's mode function. However it affects the 
evolution of $\beta_k$ as shown in Figure~\ref{betasmooth}. 
The behaviour of $|\beta_k|^2$ is modified near the reheating phase, but it coincides with the result from instantaneous reheating after a few $e$-folds. 
Hence the final value of $|\beta_k|^2$ is not affected by this modification to the Hubble evolution. This is because the post-inflation squeezing is determined by the field dynamics near the onset of the oscillations; the Hubble rates in (\ref{hubblesmooth}) and (\ref{Hubbletau}) differ for a few $e$-folds around reheating, however they coincide by the time the oscillations begin.

\begin{figure}[!ht]
    \centering
    \includegraphics[scale=.6]{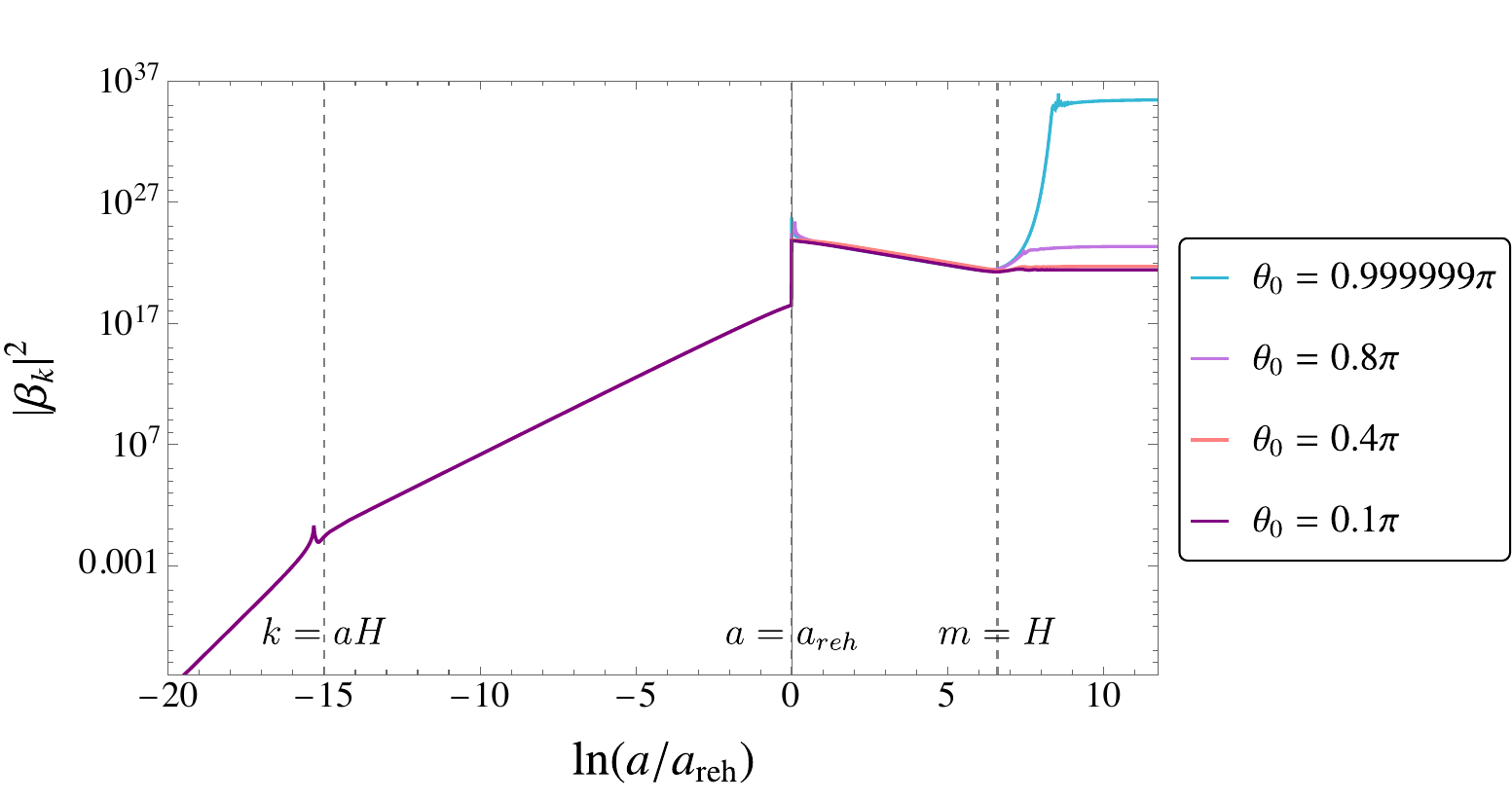}
    \caption{Time evolution of $|\beta_k|^2$ in a quasi de Sitter inflationary background, for different initial background field values. The parameters used are $m_\varphi = 10^8\,\text{GeV}$, $m=10^2\,\text{GeV}$ and $k/a_\ro{reh}=10^2\,\text{GeV}$.}
    \label{betainflaton}
\end{figure}

\subsection{Single Field Inflation}
\label{Hubblephi}

We now study a quasi de Sitter inflationary background, by considering a single-field inflation model with a quadratic potential
\begin{equation}
 U=\frac{1}{2}\, m^2_\varphi\, \varphi^2.
\end{equation}
We assume that the background dynamics is completely determined by the inflaton, with the axion field being only a spectator field. We also suppose instantaneous reheating, followed by a period of radiation domination. Therefore the Hubble rate is now given by:
\begin{equation}
\label{Hubble}
    H = \begin{cases}
        m_\varphi\, \sqrt{\frac{1}{3}-\frac{2}{3}\,\ln \left(\frac{a}{a_\ro{reh}}\right)} & \text{inflation}\\
        \frac{m_\varphi}{\sqrt{3}}\, \left( \frac{a}{a_\ro{reh}} \right)^{-2} & \text{RD}
    \end{cases}
\end{equation}

In contrast to the exact dS background invoked in the main text,
now the Hubble rate varies in time during inflation, with the inflation scale set by the inflaton mass. 
Upon computing the evolution of the axion fluctuations in the above cosmological background, we have chosen $m_\varphi = 10^{8}\, \text{GeV}$ (hence $H_\ro{inf}\sim10^{8}\, \text{GeV}$), 
combined with $m=10^2\,\text{GeV}$ and $k/a_\ro{reh}=10^2\,\text{GeV}$;
these parameters have the same order of magnitude as those in 
Figure~\ref{betasmooth}.
We checked that the mode function present similar behaviours as in the  case of an exact dS spacetime.
The results for the time evolution of $|\beta_k|^2$ are reported in Figure~\ref{betainflaton} for different initial misalignment angles. 
The general features of the evolution of $|\beta_k|^2$ are similar to those in Figure~\ref{betasmooth} for an exact dS spacetime followed by an instantaneous reheating; $|\beta_k|^2$ approaches a constant value after the onset of the oscillations, and its numerical value depends on the initial field.

From these analyses we conclude that our main conclusions for the axion squeezing are independent of the details of the inflation and reheating models.

\bibliographystyle{JHEP}
\bibliography{ref}

\end{document}